\documentclass[manuscript,screen]{acmart}
\AtBeginDocument{%
  }

\setcopyright{none}
\acmJournal{TACO}
\acmDOI{}
\usepackage{enumitem}
\usepackage{listings}
\usepackage{xcolor}
\usepackage{siunitx}

\definecolor{codegreen}{RGB}{0,120,0}
\definecolor{codeteal}{RGB}{70,120,120}
\definecolor{codeblue}{RGB}{0,40,220}

\lstdefinelanguage{CAndPython}{
  morekeywords={
    uint64_t, uintptr_t, cspacepath_t, size_t, CONFIG_WORD_SIZE, arg,
    def, return, import, as, for, in, if, else, elif, class, self,
    struct, typedef, endif, defined,
  },
  sensitive=true,
  morecomment=[l]{//},
  morecomment=[s]{/*}{*/},
  morestring=[b]",
  morestring=[b]'
}

\begin{document}

\title{CapBuddy: Efficient Physical Memory Management for Capability-based Resource-controlling Microkernel}
\author{Zhuang Lu}
\authornote{Zhuang Lu and Guangtao Zhu contributed equally to this work. Shijun Zhao is the corresponding author.}
\email{luzhuang@iie.ac.cn}
\orcid{0009-0008-8432-5828}
\affiliation{%
  \department{State Key Laboratory of Cyberspace Security Defense}
  \institution{Institute of Information Engineering, Chinese Academy of Sciences}
  \city{Beijing}
  \country{China}}
\affiliation{%
  \department{School of Cyber Security}
  \institution{University of Chinese Academy of Sciences}
  \city{Beijing}
  \country{China}}

\author{Guangtao Zhu}
\authornotemark[1]
\email{guangtao.zhu@student.unsw.edu.au}
\orcid{0009-0009-0355-810X}
\affiliation{%
  \institution{UNSW Sydney}
  \city{Sydney}
  \state{NSW}
  \country{Australia}}

\author{Shijun Zhao}
\correspondingauthor
\email{zhaoshijun@iie.ac.cn}
\orcid{0009-0000-6801-508X}
\affiliation{%
  \department{State Key Laboratory of Cyberspace Security Defense}
  \institution{Institute of Information Engineering, Chinese Academy of Sciences}
  \city{Beijing}
  \country{China}}

\author{Qianying Zhang}
\email{qyzhang@cnu.edu.cn}
\orcid{0000-0002-3246-9474}
\affiliation{%
  \department{College of Information Engineering}
  \institution{Capital Normal University}
  \city{Beijing}
  \country{China}}

\author{Lixin Chen}
\email{chenlixin@iie.ac.cn}
\orcid{0009-0004-9643-3309}
\affiliation{%
  \department{State Key Laboratory of Cyberspace Security Defense}
  \institution{Institute of Information Engineering, Chinese Academy of Sciences}
  \city{Beijing}
  \country{China}}

\author{Zekun Xu}
\email{2211002003@cnu.edu.cn}
\orcid{0009-0004-8918-4375}
\affiliation{%
  \department{College of Information Engineering}
  \institution{Capital Normal University}
  \city{Beijing}
  \country{China}}

\author{Wu Luo}
\email{luowu@iie.ac.cn}
\orcid{0009-0005-0080-8258}
\affiliation{%
  \department{State Key Laboratory of Cyberspace Security Defense}
  \institution{Institute of Information Engineering, Chinese Academy of Sciences}
  \city{Beijing}
  \country{China}}

\author{Rui Hou}
\email{hourui@iie.ac.cn}
\orcid{0000-0002-9215-7632}
\affiliation{%
  \department{State Key Laboratory of Cyberspace Security Defense}
  \institution{Institute of Information Engineering, Chinese Academy of Sciences}
  \city{Beijing}
  \country{China}}

\renewcommand{\shortauthors}{Lu et al.}

\begin{abstract}
  Capability-based resource-controlling microkernels securely multiplex hardware by exposing low-level resource abstractions guarded by capability systems.
  Such a design helps provide an application-specific resource management policy to achieve high performance without sacrificing security. However, the misuse of capability-protected secure abstractions can introduce performance overhead in user-level resource management, affecting the runtime performance of dynamic applications in emerging scenarios. In this paper, we focus on the secure abstraction of physical memory in microkernels with capability-based resource management delegation to investigate an appropriate design for dynamic memory management. We propose CapBuddy, an approach to building an efficient buddy system on capability-protected secure abstractions of memory, to bypass the performance overhead of the capability system by offloading functionalities that can run offline from the critical path of serving a memory request. We implemented a prototype system on the seL4 library OS, and the evaluation results show that our approach improves the performance of the seL4 user-level buddy system by 20 to 54.5 times and achieves the same order of magnitude as Linux. Further burst experiments evaluate CapBuddy's scalability and show that it maintains low median allocation latency as the number of concurrent clients increases.
\end{abstract}





\maketitle

\section{Introduction}

Over three generations~\cite{mmimpact93,ukernel95,sel4ten13}, microkernels from L3 to seL4 have steadily evolved toward a primary goal: securely multiplexing hardware resources among different programs~\cite{sel4ten13,cherios21,spliter25}. To achieve this, they provide simple, low-level abstractions that closely match the hardware while protecting them through a capability system for security purposes~\cite{keykos92,eros99,l4cap09}. Operating systems (OS) built on top of the microkernel can safely use these abstractions to create a library OS (LibOS) that provides application-specific resource-management policies. 
%Such design ideas reflect the spirit of Exokernel~\cite{exokernel95, exokernel97}, but extend it further by adding strong protection guarantees through capability systems. 
The feasibility of such a design idea was justified years ago by Exokernel~\cite{exokernel95, exokernel97}, but microkernels extend it further by adding strong protection guarantees through capability systems. 
Consequently, microkernels are now widely used in systems where safety and security are critical~\cite{lionsos25}, as they provide strong isolation and a small trusted computing base (TCB).

Meanwhile, recent works on microkernel-based systems such as \emph{M3}~\cite{m3codesign16}, \emph{HongMeng}~\cite{hm24}, \emph{MettEagle}~\cite{metteagle25}, and \emph{Split-container}~\cite{spliter25} show that microkernels are evolving towards new scenarios (e.g., mobile devices, heterogeneous devices, and cloud computing). 
A key merit of the above works is that they demonstrate the feasibility of a microkernel-based approach for achieving performance, dynamism, and scalability. However, these systems fall short of the policy-free ideal of microkernels, as they either retain resource-management policy in the kernel or fail to securely delegate resource management to the user level. Therefore, we argue that the aforementioned works leave a central question unanswered---\textbf{Can we still achieve competitive performance with a policy-free microkernel without sacrificing security?}---in particular, when the kernel delegates hardware resources such as memory to the user level via capability-protected kernel abstractions?

The above question matters because keeping memory management inside the kernel is risky unless the approach can be formally justified or verified~\cite{sel4ten13, Nelson_SZJBTW_17}. As evidence, researchers have found covert channels in in-kernel memory management from one of the above cases~\cite{Peter_PVNDS_16} (i.e., L4Re~\cite{Hartig_HLSW_97} under MettEagle and M3). Fundamentally, the core tension in a policy-laden microkernel is that efficiency gains may come at the expense of the very isolation that gives microkernels their security advantages. In contrast, once memory management is delegated to the user level, such threats are confined to user-level components, reducing the chance of breaking isolation. We refer to such microkernels as ``capability-based resource-controlling microkernels''.

The remaining challenge is performance, i.e., whether user-level physical-memory management can remain efficient. Fundamentally, there is a trade-off between performance and protecting hardware resources with a capability system, as each access to a capability-protected resource inherently incurs a security check by the capability system~\cite{hm24}. The cost compounds when the capability system resides in the kernel~\cite{sel4fm09,keykos92}, as each check further requires a system call with two CPU mode switches. We further argue that existing work on user-level memory management is incomplete, as it struggles to find a viable solution for capability-protected resources. Some systems provide only ad hoc or inefficient user-level policies (e.g., CantripOS~\cite{google2025cantripos}, Barrelfish OS~\cite{multikernel09, barrelbg21}, and seL4 LibOS~\cite{seL42025libos}) while others avoid dynamic allocation altogether by statically draining memory at startup (e.g., Genode~\cite{Feske:genode}, LionsOS~\cite{lionsos25}).

In this paper, we aim to explore efficient user-level methods for managing capability-protected resources on microkernels. We focus on physical memory,
as they are crucial for supporting dynamic OS mechanisms such as copy-on-write~\cite{mufork25}, on-demand process spawn~\cite{ukunfit24, ukunfit25, mufork25}, and anonymous mappings~\cite{hm24}, which are crucial in the emerging scenarios. 
Given the aims, we propose to build a high-performance, user-level buddy system on the seL4 LibOS~\cite{seL42025libos}. We are motivated by the following reasons:
\begin{itemize}
   \item The buddy system is a widely used approach for physical memory management, making it a natural first step for us to explore high-performance user-level management of capability-protected resources~\cite{knuth97, unikraft21}.
   \item seL4 is a policy-free microkernel with capability-based resource control~\cite{sel4pm08, sel4ten13}.
   \item LibOS is practical for providing application-specific policies for resource management to improve runtime performance~\cite{exokernel95, Schatzberg_CDKA_16, unikraft21}, making it a good match for microkernels whose resource management is delegated to the user level.
\end{itemize}

To investigate how much performance impact a capability system can have on memory management, we analyze the canonical dynamic memory allocator from the seL4 LibOS~\cite{seL42025libos}, which is a long-standing buddy system that has served the seL4 ecosystem for over twelve years.
The result shows that the buddy system in seL4 LibOS is 101.8 times slower than its counterparts in other LibOSes (e.g., Unikraft~\cite{unikraft21}) and 33.4 times slower than Linux. We also implemented a capability system for the Linux buddy system to remove the noise of kernel architecture. Performance analysis indicates that it likewise becomes 2.3 times slower due to the introduction of a capability system.

However, it is unjustified that memory management will perform poorly with a capability system inherently. We argue that the poor performance stems from a flawed design that results from misusing the secure abstractions of microkernels. Misuse can introduce a long call chain for applications to access resources. In the context of seL4 LibOS, such misuse leads to a design that places ``the cost of the capability system''~\cite{sel4mm-phd10, barrelbg21}
%and "the overhead of capability bookkeeping"~\cite{barrelbg21}
on the critical path of physical memory management. The cost increases the response latency for applications served by the memory allocator, leading to a performance bottleneck.

We further argue that an appropriate design can bypass the above bottleneck. To prove this claim, we propose CapBuddy, an approach to building an efficient buddy system on microkernels with capability-based resource management delegation. CapBuddy provides a user-level abstraction of memory for enforcing memory management policies. Building around the new abstraction, CapBuddy breaks a memory allocator into two halves via adopting a ``split-design'', which offloads the functionalities that can run offline from the critical path of serving a memory request. It intercepts state updates at the user level whenever possible, without passing them to the kernel and performing capability bookkeeping asynchronously. %The state interception is made through the implementation of a user-level abstraction of memory that builds over the kernel abstraction of physical memory, which makes it easier for the user level to enforce memory management policies.

To validate our design, we implemented a CapBuddy prototype on seL4 LibOS and compared it against baseline systems (e.g., Linux). The results show that our prototype achieves performance in the same order of magnitude as Linux. To demonstrate the universality of CapBuddy, we implemented two buddy system variants in the prototype for different purposes: a common variant (similar to Linux) and a specialized variant for optimization. The latter improves the batching behavior with an acceptable overhead of 5\% $\sim$ 10\%. The contributions are as follows.

\begin{itemize}[leftmargin=0.4cm, rightmargin=0.2cm, itemsep=0pt, topsep=0pt, parsep=0pt, partopsep=0pt]
    \item We explain the reasons why and how the misuse of secure abstractions in microkernels leads to poor performance in dynamic memory management.
    % \item We propose an approach for dynamic physical memory management that enables microkernels with capability-based resource management delegation to adapt to new scenarios without sacrificing performance.
    \item We propose an approach for dynamic user-level physical memory management on capability-based resource-controlling microkernels, which opens up the possibility of supporting efficient user-level policies in emerging scenarios while still providing competitive performance.
    \item We implement a high-performance buddy system prototype for seL4 and its ecosystem, improving the seL4 LibOS buddy system’s performance by 20 to 54.5 times and matching that of the Linux buddy system.
\end{itemize}
\section{Background}
\subsection{Capability System and Its Variant in seL4}\label{bg:sel4-cap-system}

Access control is a mechanism that defines \emph{who} (subject) can access \emph{what} (object) in \emph{which} ways~\cite{Lampson_71}. Microkernels usually implement access control through a \emph{capability system}~\cite{keykos92,eros99,l4cap09}, which is a system that associates each subject (user-level programs) with a list of unforgeable tokens that contain the permissions to access different objects (kernel abstractions). A \emph{capability} is an unforgeable token that contains the permissions required to access an object. To access a capability-protected object, a program must present the capability associated with that object and pass the security checks enforced by the capability system. In this paper, we refer to the operations required for a subject to pass the security checks of the capability system as the \emph{cost of capability}.

seL4 uses a capability system to protect ``kernel objects'', which refer to a system resource that the kernel maintains and exposes to user-level programs (i.e., the kernel abstractions). The primary usage of these capability-protected resources is enabling systems construction on seL4. Different user-level programs can access identical kernel objects through capabilities with different permissions. seL4 manages the capabilities of user-level programs through \emph{CSpace}, an abstraction of a hierarchical namespace for organizing capabilities. Each CSpace consists of \emph{CNodes}, each of which is an array of slots while each slot stores a capability~\cite{Boyton_09}. On seL4, the capability system functions within the kernel, which means the user-level programs should access capabilities via syscalls by providing a pointer-like reference (called \emph{CPtr}) that points to a slot (a.k.a, \emph{cslot}) in CSpace.

In seL4, all kernel objects derive from the \emph{Untyped} kernel object, which represents unused memory. From the perspective of a user-level program, an \emph{Untyped} object is the kernel abstraction of physical memory. A user-level program can use the untyped memory to create new kernel objects for different purposes, e.g., threads, pagetable structures, or a new \emph{Untyped} object. Upon object creation, the user-level program invokes a \emph{retype} syscall that instructs the kernel to convert a part of an \emph{Untyped} object into a ``typed'' kernel object (i.e., an object whose type is specified). Each retype operation initializes the new objects from an \emph{Untyped} object and generates a new capability for accessing the derived objects accordingly.

\subsection{The Evolution of Kernel Abstractions of Physical Memory on Microkernels}\label{bg:evolution}

% \begingroup
% \color{purple}
In this section, we review the evolution of the kernel abstraction of physical memory across three generations of microkernels, aiming to understand the trade-offs. We then use seL4 as a case study to describe how capability-based resource-controlling microkernels abstract memory.
% \endgroup

Mach~\cite{Young_TRGECBBB_87} represents the first generation of microkernels, which do not expose any abstraction of physical memory to the user level for resource management delegation. Instead, it manages physical memory inside the kernel. This design violates the policy-free principle of microkernels~\cite{nucleus70}. It also reduces flexibility and limits the ability of user-level systems to customize policies for varying applications~\cite{liedtke1996toward}.

L4~\cite{ukernel95} represents the second generation of microkernels. L4 also does not expose a physical memory abstraction to the user level, but it addresses part of Mach's issues by separating mechanism from policy~\cite{ukernel95}. L4 reserves a portion of physical memory as a kernel heap and provides mechanisms for recursively constructing address spaces and creating page mappings~\cite{ukernel95, weigand2004generalized}. User-level systems then manage the remaining physical memory based on these mechanisms. This design improves flexibility. However, it requires the kernel to maintain a mapping database, which consumes a large amount of kernel-heap memory and exposes the kernel to denial-of-service (DoS) attacks~\cite{sel4ten13}.

seL4, the world's first formally-verified microkernel~\cite{sel4comp14}, represents the third generation of microkernels~\cite{sel4ten13}. Building on the experience of the previous two generations, seL4 delegates physical-memory management to user space via \emph{Untyped} objects. The sole purpose of \emph{Untyped} is creating kernel objects from unused memory, i.e., treats all memory as the kernel heap, but it is not implied that all the unused memory is only for kernel use. seL4 provides another abstraction of physical memory over \emph{Untyped} called \emph{Frame}, which represents a piece of memory that is mappable to an address space. Other than \emph{Untyped} and \emph{Frame}, seL4 provides abstractions for memory that are mapped as pagetable structures for MMU/IOMMU, which are architecture-dependent. Such a design allows user space to implement application-specific memory management policies~\cite{sel4ten13}, reduces the risk of DoS attacks against the kernel~\cite{sel4ten13}, as all physical memory are seen as the heap, and makes the life cycle of each kernel object traceable through capability-based access control over physical memory, which helps the formal verification of the kernel~\cite{sel4ten13, sel4fm09}. The idea of delegating memory management out of the kernel via abstractions has influenced the design of other modern OSes and kernels, e.g., Barrelfish OS~\cite{multikernel09}, CheriOS nanokernel~\cite{cherios21}, and Asterinas~\cite{Peng_TZLCJXWXZLYZ_25, cortenMM25}.

\subsection{LibOS and Buddy System}\label{bg:libos}

A library OS (LibOS)~\cite{exokernel95}, a.k.a, \emph{unikernel}~\cite{unikernel13}, places OS services and the application in the same address space. Applications can access the services via function calls, avoiding mode-switch overhead. LibOSes are famous for providing application-specific policies through customized OS services~\cite{unikraft21}.

The buddy system~\cite{knuth97} is a popular method for physical memory management in OSes~\cite{unikraft21, Wrenger_RHDL_23}. It serves requests with varying size requirements by partitioning memory into blocks whose sizes are powers of two (e.g., 4KB, 8KB, etc.). Specifically, when a user requests a block of memory of a given size, the buddy system locates the smallest available block that is at least the requested size, then recursively splits it into pairs of equal-sized \emph{buddy} blocks until the size matches the request. When freeing a memory block, the system marks it as free and checks its buddy block. If both blocks are free, it merges them into a larger block and continues merging recursively.

%In practice, the recursive operations for splitting and merging are on the critical path for serving a memory request. If the latency of each split and merge operation is high, these operations will become the system's performance bottleneck. Different implementations may result in different performance. For example, the Linux buddy system merges memory blocks by manipulating physical addresses and flipping certain address bits during buddy block locating, which makes merging efficient~\cite{love2010linux}. In contrast, the buddy system in seL4 LibOS currently places multiple system calls on the call chain for split and merge operations, resulting in poor performance.

\subsection{Case Study: seL4 LibOS and Its Resource Management Framework}\label{bg:seL4-libos}

We use seL4 LibOS~\cite{seL42025libos} as an example to show how a user-level system manages physical memory on top of microkernels with capability-based resource management delegation. We will begin with how memory is initialized as \emph{Untyped}, then introduce how the allocators within LibOS work.

\textcolor{black}{\paragraph{\textbf{Initialization.}} At boot time, the kernel initializes all available memory and has it mapped into the kernel address space. It then generates \emph{Untyped} capabilities for the available memory, delegates all resources via capabilities as ``bootstrap information'' (\texttt{BootInfo}) to the first user-level process, i.e., the init-task, and schedules the init-task for user-level initialization.}

\textcolor{black}{The init-task is a process that parses the \texttt{BootInfo} and initializes other user-level components. It can set up new CSpace for user-level components, and dispatches the \emph{Untypeds} by putting their capabilities into the CSpaces of other components. Components who has been granted the capabilities to access the \emph{Untypeds} and CSpaces can manage the memory. Such components are usually a trusted pager, a centralized memory manager, or a LibOS application that controls the memory itself. Once initialized, the components will use seL4 LibOS to bootstrap and handle runtime memory requests.}

\paragraph{\textbf{Allocator Internals.}} In seL4 LibOS, the client program and the LibOS reside in the same address space, and a user-level memory allocator in the LibOS handles physical memory requests. A client program can be a pager, a virtual machine monitor (VMM), or even an application. The allocator consists of three components: a metadata manager, an \emph{Untyped} manager, and a capability bookkeeper.

The \emph{Untyped} manager records the size and state of available \emph{Untyped} objects, and selects a suitable \emph{Untyped} object when an application requests memory. The capability bookkeeper uses a bitmap to record the availability of the in-kernel \emph{cslots}, which store the capabilities of kernel objects and the \emph{Untypeds}. The metadata manager provides metadata allocation support. It controls the heap of the LibOS and receives metadata bookkeeping requests from the other parts of the allocator, i.e., the \emph{Untyped} manager and the capability bookkeeper. 

\autoref{fig:application_allocator_interaction} shows how an application on seL4 LibOS interacts with the allocator through a UNIX \texttt{mmap}-like request. In this example, the application requests a one-page, read-only, and anonymous mapping at a given virtual address. When the allocator receives the request, it explicitly allocates a \emph{Frame}, a kernel object representing mappable physical memory, as the backing memory. The allocator then returns a pointer to a \emph{cslot} that stores the \emph{Frame} capability, which will be passed to the mapping syscall to install the mapping in the target address space (also referenced by a capability pointer).

\begin{figure}[htb]
\centering
\begin{lstlisting}
vaddr = arg(0)                                            # virtual address to map
cptr_vspace = arg(1)                                      # pagetable capability pointer
cptr_frame = sel4_libos_allocator_alloc(KOBJ_TYPE_FRAME)  # allocate a Frame kernel object
# mmap(vaddr, PAGE_SIZE, PROT_READ, MAP_ANONYMOUS | MAP_PRIVATE, -1, 0)
err = seL4_Syscall_Map(cptr_vspace, cptr_frame, vaddr, PROT_READ, MAP_ANONYMOUS | MAP_PRIVATE, ...)
\end{lstlisting}
% /* (given user args and variables:
%  *  -> a pagetable cptr: given as 'cptr_vspace'
%  *  -> a virtual address to map: 'vaddr')
%  */
% uint64_t cptr_vspace = arg(0);
% uint64_t vaddr = arg(1);
% /* mmap(vaddr, PAGE_SIZE, PROT_READ,
%  *      MAP_ANONYMOUS | MAP_PRIVATE, -1, 0)
%  */
% uint64_t cptr_frame = 
%     sel4_libos_allocator_alloc(KOBJ_TYPE_FRAME);
% /* once allocated, use the frame cptr and map it
%  */
% uint64_t err = seL4_syscall_create_mapping(
%     cptr_vspace, cptr_frame,
%     vaddr, PROT_READ, MAP_ANONYMOUS | MAP_PRIVATE, ...
% );
\caption{Pseudocode of how an application requests a \emph{Frame} object
from seL4 LibOS and maps it into an address space.}
\label{fig:application_allocator_interaction}
\end{figure}

\begin{figure}[htb]
\centering
\begin{lstlisting}
def sel4_libos_allocator_alloc(obj_type: int) -> int:      # type is provided by the application
    cptr = capability_bookkeeper_slot_pin()                # reserve a CSpace slot for the new capability
    cookie = untyped_manager_alloc(cptr, obj_type)         # create the object and place its cap at cptr
    allocator_internal_record(cookie, cptr, obj_type)      # record allocator metadata for the created object
    return cptr                                            # return the CPtr used to invoke the new object
\end{lstlisting}
% /* type is provided by the application */
% uint64_t sel4_libos_allocator_alloc(uint64_t type)
% {
%     /* reserve a CSpace slot for the new capability */
%     uint64_t cptr = capability_bookkeeper_slot_pin();

%     /* create the object and place its cap at cptr */
%     uint64_t cookie = untyped_manager_alloc(cptr, type);

%     /* record allocator metadata for the created object */
%     (void) allocator_internal_record(cookie, cptr, type);

%     /* return the CPtr used to invoke the new object */
%     return cptr;
% }
\caption{Pseudo code of how seL4 LibOS handles request.}
\label{fig:allocator_internal_interactions}
\end{figure}

We also present the workflow of how the allocator responds to an allocation request in~\autoref{fig:allocator_internal_interactions}. First, the allocator instructs the capability bookkeeper and pins a slot for the requested object. The \emph{Untyped} manager then selects a suitable \emph{Untyped} object regarding the request, and asks the kernel to perform a \emph{retype} syscall, which creates the target object and stores its capability into the pinned slot. The process of de-allocation is in reverse order of allocation, i.e., the \emph{Untyped} manager instructs the kernel to delete the kernel object first, the \emph{Untyped} manager then restores the state of the \emph{Untyped} object, and the bookkeeper releases the pinned slot once the above process finishes.
\section{Quantitative Analysis and Insights}\label{a:analysis-and-insights}

This section analyzes the impact of capability systems on memory management performance and presents our design insights.
We analyze from two aspects. First, we quantify the performance overhead of allocation and deallocation operations in the buddy system of seL4 LibOS (hereafter referred to as seL4 Buddy), showing how capability systems affect the basic operations of a user-level buddy system. Second, we provide a breakdown analysis to the overhead to investigate the root cause. We hypothesize that the impact of capability systems on buddy systems is not limited to system calls; it also introduces a longer call chain into the state updating operations. To validate the hypothesis, we implement a capability-protected buddy-system prototype in the Linux kernel. This prototype converts the capability invocations that are performed via seL4 syscalls into function calls. 
We then propose some insights, which help us design a buddy system suitable for capability-protected physical memory.
%\textcolor{purple}{We then propose some insights about why the design of seL4 Buddy needs a change, and what could have been done better within the design space.}

% \subsection{Capability-System Overhead in seL4 Buddy}

\subsection{Performance Bottleneck in seL4 Buddy}\label{a:sel4-buddy-bottleneck}
\begin{figure*}[t]
  \centering
  \includegraphics[width=\textwidth]{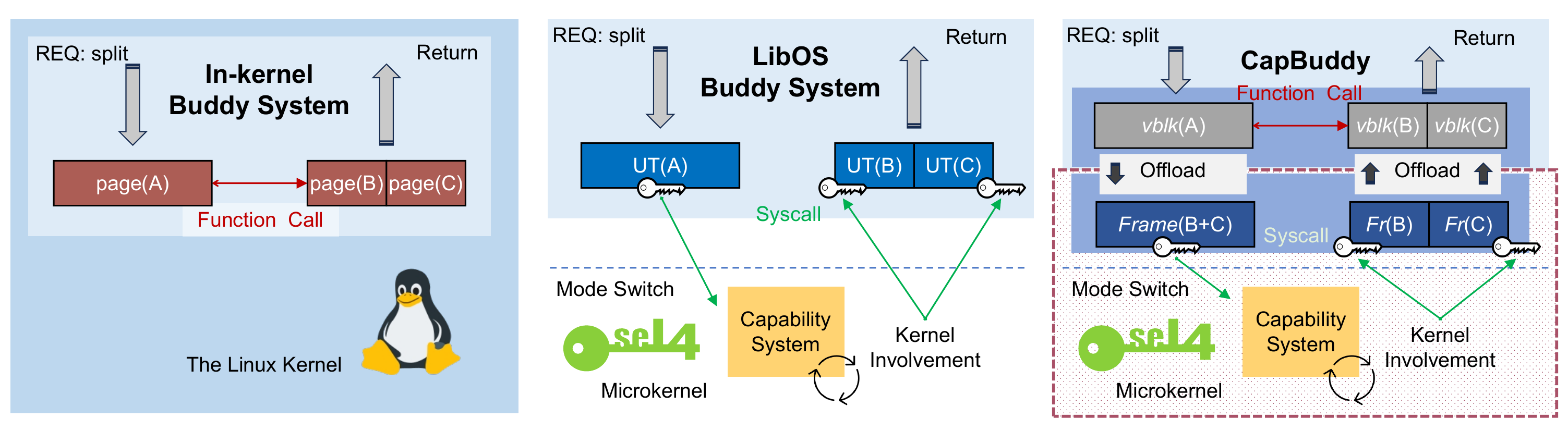}
  %\captionsetup{skip=4pt}
  \caption{Different buddy systems among the Linux kernel, seL4 LibOS, and CapBuddy over seL4 LibOS. Components within the dark-red, dotted box are operations for materializing the \emph{vblks}, which are not on the critical path.}
  \label{fig:linux-vs-libos-vs-capbuddy}
\end{figure*}

By comparing seL4 Buddy with the Linux buddy system, we found that the performance bottleneck of seL4 Buddy mainly stems from its reliance on a physical memory abstraction that requires costly state updates (\autoref{fig:linux-vs-libos-vs-capbuddy}). First, seL4 Buddy and the Linux buddy system adopt a similar design approach, as both implement the buddy algorithm (i.e., split and merge operations) based on their respective physical memory abstractions. The difference is that seL4 Buddy relies on the \emph{Untyped} object, whereas the Linux buddy system is based on \texttt{struct page}\footnote{The definition of \texttt{struct page} is available in \url{https://github.com/torvalds/linux/blob/master/include/linux/mm_types.h}.}. Furthermore, we discovered that seL4 Buddy's physical memory abstraction causes the split/merge operations to trigger a long call chain within the seL4 LibOS, which requires not only syscalls but also complex capability bookkeeping. Specifically, the long call chain of seL4 Buddy consists of the following three stages. In contrast, the memory abstraction underlying the Linux buddy system allows it to execute the buddy algorithm using only stage 1, requiring no syscalls but only in-kernel function calls
\begin{enumerate}
    \item \textbf{Updating block metadata.} The memory allocator updates the metadata describing the block whenever it is split or merged.
    \item \textbf{User-level capability bookkeeping.} The allocator allocates or frees a \emph{cslot} for recording a protected block and updates accounting data.  Such processes can involve the overhead of searching the target \emph{cslot}. %(searching overhead evolved, at best $\mathcal{O}(logn)$).
    \item \textbf{Capability Invocations.} Every basic operation can bind with an seL4 syscall to split or merge the blocks, which leads to two switches between machine modes, one security check at each kernel entry, and an internal state update of the capability system. 
\end{enumerate}

\begin{figure}[htbp]   
    \centering
    \includegraphics[width=0.6\columnwidth]{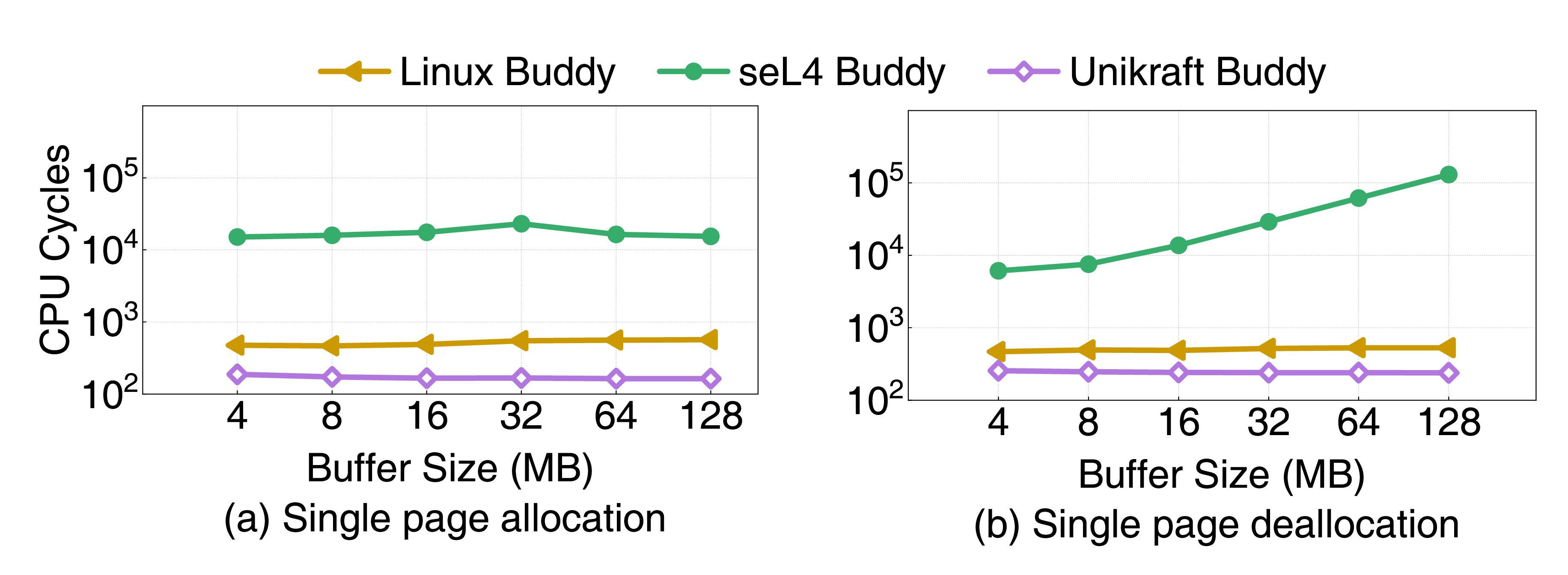}
    \caption{Single-page performance comparison between seL4, Linux, and Unikraft buddy systems.}
    %\captionsetup{skip=4pt}
    \label{fig:problem_analysis}
\end{figure}

From the above facts, we realized a design flaw of seL4 Buddy is that it builds a synchronous model of buddy system on abstractions that are costly to update. To validate our thoughts, we compared seL4 Buddy's performance with other baselines (i.e., Linux and Unikraft~\cite{unikraft21}). The results in \autoref{fig:problem_analysis} show that seL4 LibOS underperforms other systems in both allocation and deallocation by orders of magnitude; see \autoref{eval} for details.

\subsubsection{\textbf{The Cost of Capability Invocations}}

One major overhead in seL4 Buddy stems from operations that access the in-kernel capability systems, which lead to the overhead of mode switches and capability security checks.

Suppose we are allocating a 4 KB \emph{Frame} out of a 4 MB \emph{Untyped} via seL4 Buddy. We consider the worst-case splitting scenario, where no smaller suitable \emph{Untyped} is available and splitting must start from 4 MB. In this case, it will need 11 system calls in total: 10 to split the \emph{Untyped} to new \emph{Untypeds} of 4 KB, and 1 to retype a 4 KB \emph{Untyped} into a \emph{Frame} for actual mapping. Each split halves the size of the \emph{Untyped}. Since 4 MB is $2^{10}$ times 4 KB, 10 split operations are required along the allocation path of the target \emph{Frame}. To quantify the overhead, we measured the above operations on an ARM platform (Cortex-A53, 1.4 GHz), where the average cost of one \emph{retype} call is $1,396$ cycles (breakdown see \autoref{t:buddy-data-access-breakdown}). In total, the above split operations take $16,625$ cycles. By contrast, performing the same split in Linux Buddy requires 137 cycles on average and 1,596 cycles in total.

\begin{table}[h]
  \centering
  \caption{The breakdown of the average cost of splitting a block (cycles). ``Kernel Path'' includes the cost of capability checks, metadata update triggered by the split operation, and scheduler involvement in seL4.}
  \label{t:buddy-data-access-breakdown}
  \begin{tabular}{l | r r r }
    \bf Variant & \bf Mode Switch& \bf Kernel Path & \bf Ave\\
    \hline
    \bf Linux       & - & -  & 137 \\
    \bf seL4 LibOS  & 295 & 1,101 & 1,396 \\
    \hline
  \end{tabular}
\end{table}

\subsubsection{\textbf{User-level Capability Bookkeeping}}\label{sss:user-level-cap-bookkeeping}

Complex capability bookkeeping is another significant source of overhead in seL4 Buddy, which relies on a capability bookkeeper to record the availability of \emph{cslots}. Every memory request requires allocating/reclaiming slots and metadata. Concretely, when a user requests one block, the allocator must search for a slot in the CSpace, allocate memory for metadata, and initiate the metadata to track the slot’s state. We measured the average cost of allocating a \emph{cslot} on the same platform as above, the latency is 568 cycles. Note that the reason why seL4 Buddy spends hundreds of cycles searching for \emph{cslots} is the lack of optimization, as it adopts a design that uses bitmaps plus linear scan searching without batching as the bookkeeping policy for capabilities.

\subsection{Validating the Long-Call-Chain Hypothesis}

To evaluate the performance impact of the long call chain without the mode-switch overhead of system calls, we built a capability system prototype in the Linux kernel for its buddy system. This prototype follows seL4 Buddy’s design principles, introducing a long call chain for each basic operation of the buddy system in Linux (i.e., adding the length of the critical path for serving a request). This experiment isolates and validates the performance impact of the long call chain itself.

\begin{figure}[t]   
    \centering
    \includegraphics[width=0.6\columnwidth]{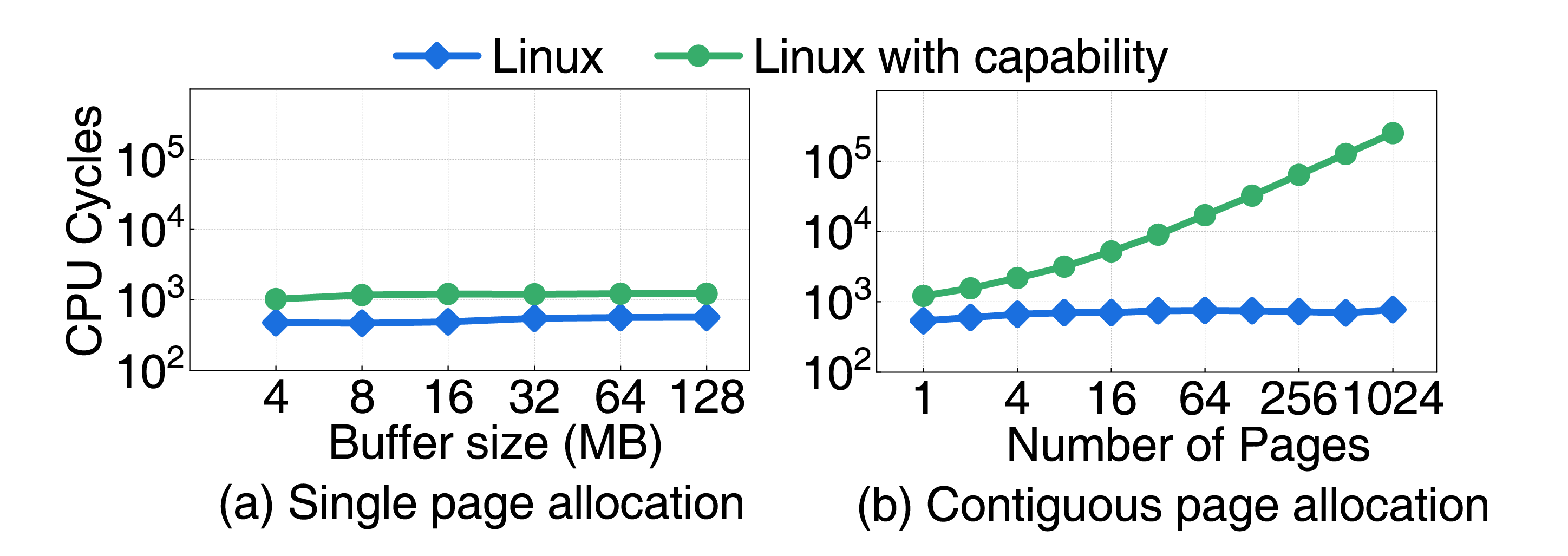}
    \caption{Performance impact of the capability system on Linux Buddy.
    % for single-page and contiguous physical memory allocation.
    }
    \label{fig:linux_cap}
\end{figure}

The specific design includes: (1) maintaining a simplified capability derivation tree~\cite{Boyton_09} for each block of physical memory, (2) simulating seL4 LibOS’s user-level capability bookkeeping model, and (3) capability invocations as function calls that perform security checks. 
%We conduct a vertical comparison with the default Linux Buddy without a capability system. 
The evaluation results (\autoref{fig:linux_cap}) show that the introduction of a capability system likewise makes the Linux buddy system 2.3 times slower on single-page allocation, while the cost increases sharply for batch allocations.

\subsection{Insights}\label{subsec:capbuddy-abstraction-design-space}

\paragraph{\textbf{Insight 1: Specific Strategy for Frame objects.}}
As described in \autoref{bg:evolution}, there is a classification about seL4 kernel abstractions of memory: \emph{Untyped} helps create other kernel objects from unused memory, which treats all memory as kernel heap. \emph{Frame} builds on \emph{Untyped} and makes a memory mappable to an address space. 
%\emph{PageTable}, which also builds on \emph{Untyped}, is for memory that maps in the pagetable structure.

seL4 Buddy serves requests of varying sizes for all kernel objects, credited by the features of the \emph{Untyped} object. 
However, we argue that a microkernel's buddy system is best suited to provide only page-sized mappable memory (i.e., \emph{Frame}), due to the following reasons. First, the distribution of the sizes of seL4 kernel objects is highly skewed and clustered around a small number of size classes and page-size classes, and among page-sized objects, \emph{Frame} objects account for the vast majority. Many architecture-independent objects~\cite{sel4manual13} are small, falling in the 16 $\sim$ 128 Bytes range, while the rest are dominated by objects of page-sized at 4 KB (e.g., \emph{Frame}, \emph{vcpu}, pagetable structures)~\footnote{Some outliers are \emph{CNode} (varying size), \emph{Untypeds} (varying size), root pagetable (8 KB, ARM64), hugepages (2 MB or 1 GB, ARM64), and \emph{thread} (2 KB).}. We analyzed two seL4-based systems (i.e., a Windows VM based on seL4 Microkit~\cite{seL42025microkit} and Google CantripOS~\cite{google2025cantripos}) and found that \emph{Frame} objects account for the vast majority of all objects (98\% and 69\%, respectively) (see \autoref{tab:microkit-cantrip-obj}). Second, related research~\cite{sel4mm-phd10,unikraft21} shows that buddy systems are inherently inefficient when managing a heap where requests of varying, small sizes dominate (usually much smaller than page-size), compared to other general-purpose allocators (e.g., minimalloc, tinyalloc).

\paragraph{\textbf{Insight 2: Buddy system operations on physical memory (or memory abstractions) can be modeled as operations on array indexes.}}
A buddy system can model pages of physical memory (or memory abstractions, such as Linux's \emph{struct page}) as an array, where each element represents a page (or a memory abstraction) and the index of the element corresponds to the page's (memory abstraction's) offset from a base address. Under this view, split and merge operations can be represented as operations over array indexes. Following the same idea, if we model the capabilities of contiguous frames as an array and maintain the mapping between array indexes and CSpace slots at the user level, we can build a user-level buddy system over frame capabilities by manipulating array indexes.

\begin{table}[htbp]
    \centering
    \small
    \caption{Kernel objects counts for a seL4 Microkit-built Windows VM and CantripOS.}
    \label{tab:microkit-cantrip-obj}
    \begin{tabular}{lrrrrr}
        \toprule
        System & Frame & PageTable & VCPU & Others & Total \\
        \midrule
        seL4 Microkit & 10,619 & 147 & 1 & 86 & 10,853 \\
        CantripOS  & 245    & 20 & 0 & 89 & 354 \\
        \bottomrule
    \end{tabular}
\end{table}

\section{CapBuddy System Design}\label{sec:capbuddy-design}

We propose CapBuddy, an approach to building an efficient user-level buddy system without kernel modifications on capability-based resource-controlling microkernels. CapBuddy introduces a user-level abstraction for representing allocatable memory, called \texttt{virtual\_zone} (\emph{vzone}).

Based on insight 1 (see \autoref{subsec:capbuddy-abstraction-design-space}), \emph{vzone} manages memory at page granularity and serves as a shadow representation of page-aligned memory region. It records allocation states at the user level while maintaining the correspondence between these states and the underlying capabilities. Based on insight 2 (see \autoref{subsec:capbuddy-abstraction-design-space}),  CapBuddy enables the manipulation of the \emph{vzone} in user space to implement the buddy system, thereby allowing buddy operations to be implemented through function calls.
CapBuddy adopts a split design that triggers capability operations only when needed, which separates memory-management policies from expensive capability-protected resource provisioning. Specifically, CapBuddy divides the allocator into a front-end and a back-end. The front-end enforces buddy algorithms on \emph{vzones}, while the back-end maintains a memory pool that materializes the \emph{vzones}.

\subsection{Vzone Abstraction}\label{subsec:capbuddy-abstraction-overview}

%The definition of \emph{vzone} is presented in \autoref{fig:capbuddy_abstraction}. 
The size of each \emph{vzone} (\autoref{fig:capbuddy_abstraction}) is a multiple of the page size, which means a \emph{vzone} contains multiple page-granular blocks (for brevity, \emph{vblks}). Each \emph{vblk} has its equivalence in physical memory, while the relationship is recorded in a data field called \texttt{vblk\_sequence}, which is a list of CSpace descriptors that link the \emph{vblks} with their underlying capability-protected resources. The memory that a \emph{vzone} describes begins at \texttt{base\_physical\_address}, which is page-aligned and contiguous. 

\begin{figure}[htb]
\centering
\begin{lstlisting}
class virtual_zone:
    if CONFIG_ENABLE_BITMAP:
        if CONFIG_WORD_SIZE == 32:
            arch_data:     arch32_single_level_bitmap_t
        else:                                # 64 arch
            arch_data:     arch64_two_level_bitmap_t
    elif CONFIG_LINUX_STYLE:
        free_areas:        list[free_area]   # MAX_ORDER entries

    base_physical_address: uintptr_t         # base address of the original untyped object & vblk_sequence
    vblk_sequence:         list[cspacepath_t]# metadata for frames retyped from the original untyped
    largest_avail:         size_t            # current largest available memory region size in bits of frame number
    block_size:            size_t            # small/huge
    state:                 int               # state of the vzone

vzone_t = virtual_zone
\end{lstlisting}
\caption{Simplified definition of CapBuddy abstraction.}
\label{fig:capbuddy_abstraction}
\end{figure}

CapBuddy maintains a \emph{vzone} pool that organizes \emph{vzones} according to their \texttt{base\_physical\_address} or the first \emph{cslot} in \texttt{vblk\_sequence}. The pool maintains two invariants. First, every \emph{vzone} visible in the pool is structurally valid: its metadata must have been fully initialized before adding to the pool. Second, only \emph{vblks} marked as provisioned are exposed to the front-end as allocatable memory; such \emph{vblks} always have valid \emph{Frame} capabilities that can be returned to clients. A \emph{vblk} whose \emph{Frame} capability has been reclaimed remains tracked by its \emph{vzone}, but it is marked unavailable to the front-end until the back-end materializes it again. CapBuddy preserves these invariants by publishing and removing \emph{vzones} through synchronization mechanisms (e.g., locks or atomics). The \texttt{state} field in each \emph{vzone} records whether the \emph{vzone} is visible to pool users.

% CapBuddy maintains a \emph{vzone} pool that organizes \emph{vzones} according to their \texttt{base\_physical\_address} or first \emph{cslot} of \texttt{vblk\_sequence}. This pool functions with a simple validity invariant: every \emph{vzone} visible in the pool is fully initialized and contains valid capabilities for its \emph{vblks}. Such an invariant is achievable by updating the \emph{vzone} pool with synchronization mechanisms (e.g., locks or atomics). We also use a \texttt{state} data in \emph{vzone} to determine whether it should be visible to pool users.

\subsection{Execution Model}

CapBuddy uses a single-client, single-daemon execution model. We design CapBuddy for LibOS, so the client program and the LibOS allocators (e.g., CapBuddy) share the same address space. The front-end and the client run in the same ``client'' thread so that the front-end serves requests synchronously. The back-end runs as a daemon thread that maintains the \emph{vzone} pool. We adopt this single-client model because (1) it aligns well with LibOS because single-client applications are popular in LibOS environments~\cite{Baumann19, ukl23, mufork25}, and (2) CapBuddy focuses more on offloading capability cost than scaling the allocator across many cores.

At boot time, the LibOS runtime initializes the \emph{vzone} pool with a capacity that adjusts dynamically. The source of the memory to supply the pool is from a bootstrap allocator or external allocators, as CapBuddy does not consume all \emph{Untypeds}. The \emph{vzone} pool also has a watermark and thresholds, which are for recording the number of memory in-use and a point for the pool to ask for refill/shrink. The default capacity, watermark, and thresholds are all application-specific, blessed by the LibOS architecture. 

The front-end is the only instance that directly serves memory requests. It performs the buddy algorithm and selects an appropriate \emph{vblk} from \emph{vzones} in the pool. No resource provisioning operations are necessarily performed by the front-end unless the pool is empty. Before entering the pool, the front-end checks the current watermark. If it grows above/below a threshold, the front-end sends an asynchronous signal to the daemon and then continues the current request without waiting. This implies that the threshold for refill should be lower than the pool capacity, which promises a slack that the front-end can finish without provisioning \emph{vzones} itself.
% During each call, the front-end updates buddy metadata and returns the corresponding capabilities from \texttt{vblk\_sequence}.

The back-end daemon has high scheduling priority and remains passive until it receives a signal. Being passive means that the kernel does not schedule it unless it has been driven by events~\cite{Lyons_MAH_18}. Once woken, the daemon provisions memory in bounded batches. One refill/shrink batch is one \emph{vzone}, containing multiple \emph{vblks} (e.g., 1024 per batch). Upon refill, the daemon initializes a \emph{vzone} outside the front-end critical path, materializes its capabilities, and publishes the completed \emph{vzone} to the pool atomically. The shrink operation is similar: the daemon atomically detaches an empty \emph{vzone} from the pool, recycles it, and donates its memory to other allocators whoever is in need.

The pool slack is the key to hide provisioning latency. Since the front-end and the daemon are two different threads, the client can continue requesting existing provisioned \emph{vzones} while the daemon prepares the next \emph{vzone}. The refill threshold is set conservatively, e.g., at $50\%$ of a pool capacity, so that the remaining memory can cover the worst-case daemon refill latency under peak allocation rate.

\subsection{Resource Provisioning Design Space}

\paragraph{\textbf{Why Memory Pool?}}

\begin{figure}[htb]
  \centering
  \includegraphics[width=0.6\columnwidth]{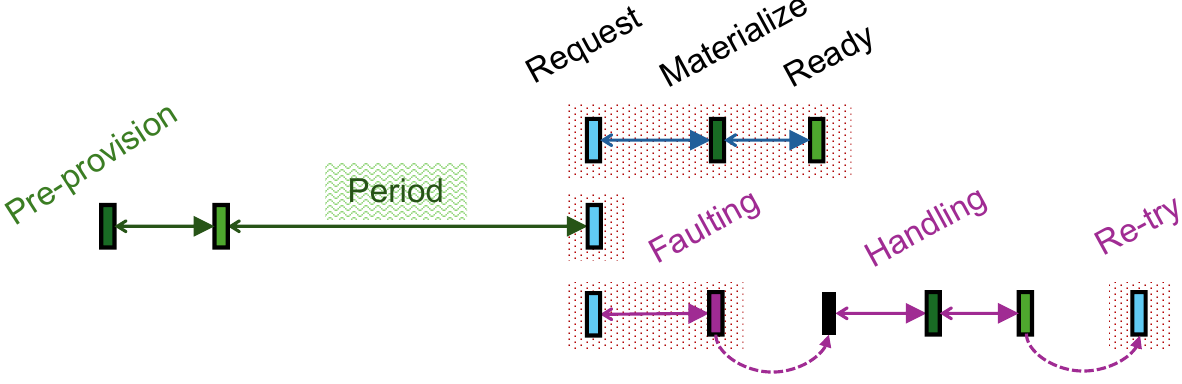}
  %\captionsetup{skip=4pt}
  \caption{Request models. Components in red dotted boxes are on the
  critical path of serving a request.}
  \label{fig:request_models}
\end{figure}

To reason about the design choice of using a memory pool, we model a client's resource request by the temporal relation among requesting, provisioning, and using a resource, as shown in \autoref{fig:request_models}.
\begin{enumerate}
    \item The request and the provisioning can be synchronous: resource provisioning happens on the request path.
    \item The request and the provisioning can be asynchronous:
    \begin{enumerate}
        \item \textbf{pre-provisioning.} Provisioning happens before the request, so the resource has already been materialized when the client uses it.
        \item \textbf{fault-driven provisioning.} The client first attempts an access without a materialized resource, and the resulting fault/event/exception triggers a handler that materializes the missing resource.
    \end{enumerate}
\end{enumerate}

Applying the above model to the buddy system in \autoref{bg:seL4-libos}, the resources under provisioning are a set of mappable memory. The request is mapping-oriented: a client requests \emph{Frames} so that it can consume them via syscalls to create mappings.

Motivated by the issue that seL4 Buddy couples state updates with expensive capability invocations, CapBuddy adopts an asynchronous provisioning model for memory resources. Being asynchronous implies more opportunities to amortize provisioning overhead, e.g., a fault-driven model with batching and interrupt coalescing~\cite{vICirqCoalescing11} can help optimize the average fault handling latency. For CapBuddy, however, the more sensible asynchronous model is to materialize \emph{Frames} before the syscalls that map them (i.e., pre-provisioning). This is because the fault-driven model does not work for seL4 syscalls, which are atomic, meaning a client cannot invoke a mapping syscall without presenting the required capability and then supply it later through a fault handler.

\paragraph{\textbf{Provisioning Granularity.}}

The basic granularity for pool management is \emph{vzone}, which means the back-end refills or shrinks the pool with memory in batches. During shrink, the daemon atomically detaches an empty \emph{vzone} from the pool, deletes all capabilities to the \emph{Frames} in that \emph{vzone}, and reclaims the \emph{cslots}. It then reclaims the corresponding \emph{Untyped} resources for use by other allocators. However, given the granularity for the front-end to consume memory is \emph{vblk} (as each \emph{Frame} behind a \emph{vblk} is the granularity for invoking syscalls), internal fragmentation within \emph{vzones} could be an issue for shrink operations, as the granularity for draining is smaller. To address this issue, CapBuddy provides a fine-grained provisioning mechanism that allows the back-end to reclaim at \emph{vblk} granularity. The back-end can pin some available \emph{vblks}, delete their \emph{Frame} capabilities and reclaim the corresponding \emph{Untyped} resources, while preserving their states within the \emph{vzone} as ``allocated''. Reclaiming the corresponding \emph{Untyped} resources must not affect other pages still in use in the \emph{vzone}. For \emph{Frames} in the \emph{vzone} created from the same \emph{Untyped}, that \emph{Untyped} can be reclaimed only after the \emph{Frame} capabilities of all corresponding \emph{vblks}, including any copies, have been deleted. Such a mechanism is similar to the idea of memory ballooning~\cite{Ballooning02}. The trade-off is that it introduces longer latencies, making \emph{vzone} the default granularity unless the fragmentation issue is severe.

To ensure both granularity for pool management work without breaking CapBuddy’s validity invariants,
% : every visible \emph{vzone} remains structurally valid, and only provisioned \emph{vblks} are exposed to the front-end as allocatable memory
 we also record the capability of \emph{Untyped} that each \emph{Frame} of \emph{vblks} is retyped from. This helps the daemon to reclaim and re-materialize these ``allocated'' \emph{vblks} safely, as their \emph{Frame} capability could have been deleted during page-granular reclaim, keeping their backing \emph{vzone} structurally valid and visible to the pool.
To prevent the front-end from misusing these ``allocated'' \emph{vblk} by requesting a deallocation on them, 
% CapBuddy provides three states to record the availability of \emph{vblks} within a visible \emph{vzone}: (1) provisioned-available, (2) provisioned-allocated, and (3) unprovisioned, which means the \emph{vblk} is occupied but unavailable to be freed.
CapBuddy records each \emph{vblk} in a visible \emph{vzone} with one of three states:
(1) \emph{provisioned-available}, meaning that the \emph{vblk} has a valid \emph{Frame} capability and can participate in front-end allocation and merge operations;
(2) \emph{provisioned-allocated}, meaning that the \emph{vblk} has a valid \emph{Frame} capability but is currently owned by a client; and
(3) \emph{unprovisioned}, meaning that the \emph{vblk} remains tracked by the \emph{vzone} but has no \emph{Frame} capability available to the front-end, so it is excluded from buddy operations until the back-end materializes it again.

\subsection{Fastpath}

The process of allocating an object from seL4 LibOS allocator was: (1) pin capability slots, (2) perform buddy algorithm, (3) allocate the object by creating new capabilities into the pinned slots, and (4) return the reference to the slots. %We refer to the entire process as a baseline path.

When using CapBuddy, however, steps (1) and (3) are not performed as long as the pool contains suitable \emph{provisioned-available} \emph{vblks}, meaning only (2) and (4) are on the critical path, which is the definition of ``fastpath''. The fastpath relies on CapBuddy's validity invariant in \autoref{subsec:capbuddy-abstraction-overview}, which ensures all allocation requests finish at the \emph{vzone} level. Once an allocation request has gone through the buddy algorithm and arrived at \emph{vzone}, the allocator returns by serving the clients with the \emph{cslots} and physical address of the allocatable \emph{vblks}. The client then uses the returned \emph{cslot} reference to map the \emph{Frame} through a seL4 system call, during which the kernel checks the corresponding capability. CapBuddy does not modify the seL4 microkernel. Operations on kernel objects still go through seL4 system calls, so these objects remain protected by the microkernel's capability mechanism.

For deallocation fastpath, the request arrives with an input of a cookie (e.g., \emph{cslot} or physical address). Before submitting this request, the client must stop accessing the page and unmap it through seL4 LibOS. It must also remove mappings established through any copied \emph{Frame} capabilities and delete those copies. The front-end uses the input to locate the \emph{vzone} to update, performs buddy algorithm, and changes the relevant states. The pool retains the existing \emph{Frame} capability for subsequent allocations, but the client must stop using the previously returned \emph{cslot} reference. On reallocation, the front-end returns the corresponding \emph{cslot} reference for a new mapping.

%The relationship is recorded by the \emph{vzone} pool according to \autoref{subsec:capbuddy-abstraction-overview}.
% To record the relationship between the deallocation input and \emph{vzones}, we use a red-black tree that organizes all available \emph{vzones} by their base \emph{vblk} indices (i.e., the base \emph{cslots}) or \texttt{base\_physical\_address}. 

No actual capability invocations or capability bookkeeping operations are involved in the fastpath. The front-end only changes the state recorded in \texttt{free\_areas} or \texttt{arch\_data}, i.e., the data field that left for building buddy algorithms on \emph{vzone}. Neither capability slots pin/unpin will happen due to capability bookkeeping, nor \texttt{retype} due to kernel object allocation. Such operations happen only when the pool is drained, while the front-end, in that case, needs to signal and wait synchronously for the daemon to refill.

\subsection{Capability Bookkeeping}

Though capability bookkeeping is largely offloaded to the daemon, a faster bookkeeper can still improve the latency for the daemon to provision memory for \emph{vblks}. 
When initializing the \emph{vzones}, the back-end uses the capability bookkeeper to pin slots for the \emph{Frame} capabilities. Given the lessons learned from \autoref{sss:user-level-cap-bookkeeping}, we provide the capability bookkeeper with batching mechanisms that help amortize the average search latency per slot. This design aligns well with how the daemon materializes \emph{vzones}, i.e., creates \emph{Frames} in batch. The capability bookkeeper now allocates contiguous \emph{cslots} in batch for each \emph{vzone} when initializing its \texttt{vblk\_sequence}. 

In seL4 Buddy, \emph{cslot} bookkeeping operations are coupled with every split/merge operation. While in CapBuddy, \emph{cslot} bookkeeping operations are coupled with \emph{vzone} materialization. The lifetime of a \emph{cslot} is then tied to a \emph{vblk}/\emph{vzone} rather than to an individual split/merge operation.
% However, there are no obvious limitations that prevent supporting dynamic capability bookkeeping, i.e., making some \emph{cslot} under unprovisioned \emph{vblks} reclaimable.

% \subsection{Pre-provisioning}

% \textcolor{purple}{\textbf{TODO}: redo this part, but yes the primary parts can stay. We also need to discuss how capability provisioning cooperates with the seL4 LibOS and the new abstractions.}

\section{CapBuddy's Prototype on seL4}

% The introduction of \emph{vzone} provides an opportunity to separate policies from \textcolor{purple}{metadata provisioning,} i.e., once a \emph{vzone} is initialized, you can assume that the allocation states updating operations would all finish at \emph{vzone}-level, instead of passing through to the in-kernel capability system on microkernels.

% To validate the hypothesis, 
We prototype CapBuddy over the seL4 microkernel and implement the \emph{vzone} abstraction using utilities provided by the seL4 LibOS (referred to as seL4-based CapBuddy). The front-end and the back-end daemon run on separate cores under seL4's SMP kernel. We base CapBuddy's prototyping work on seL4 Buddy by redirecting the requests for mappable memory from seL4 Buddy to seL4-based CapBuddy. Such a choice can (1) prevent a revamp of the entire memory management framework, and (2) reuse the framework to manage other kernel objects for convenience. We chose seL4 Buddy as a starting point because of its resource management framework, including a capability bookkeeper, a buddy system that provides a universal allocation strategy for varying kernel objects, and a minimum compatible layer that simplifies the use of capability invocations. 

%(``universal buddy system'' for convenience)

% \textcolor{purple}{\textbf{TODO}: explains the figure.}

% To present the split-design property of CapBuddy (\autoref{fig:sel4-buddy-retrofit}), the key is to implement an indexable namespace where each element is a slot of frame capabilities. We retrofit seL4 Buddy to make it function as a prototype CapBuddy (``seL4-based CapBuddy'' for convenience) by decoupling the three components with new layers in between. The indexable namespace lives in the layer between the capability bookkeeper and the universal buddy system. We then split the software stack into a front-end with an indexable namespace at the bottom and a back-end with a frame-provisioning manager on top. 
%The structure of the seL4-based CapBuddy is presented in \autoref{fig:sel4-buddy-retrofit}.

% \begin{figure}[htbp]
%  \centering
%   \includegraphics[width=\columnwidth]{figures/seL4-buddy-retrofit.pdf}
%  \caption{seL4-based CapBuddy.}
%  \label{fig:sel4-buddy-retrofit}
% \end{figure}

\subsection{Vzone Provisioning}

We use a red-black tree to organize the \emph{vzones} in seL4-based CapBuddy. For synchronization, we now use a coarse-grained lock that ensures the atomicity of state updates from the front-end and the daemon. Such an implementation can be replaced when a multi-client, multi-daemon model is required. To materialize the \emph{vzones}, seL4-based CapBuddy provides the capability bookkeeper with an interface to create and initialize a region in the CSpace for initializing \emph{vblk\_sequence}. 

\subsection{Front-end Variants.}
% Instead of relying on the universal buddy system to provide a ``buddy system'', 
With the help of \emph{vzone}, seL4-based CapBuddy provides the system designer with the ability to implement their own variants of buddy systems or algorithms over \emph{vzone}. The variants build on the policy-oriented data structure (i.e., \texttt{arch\_data} and \texttt{free\_areas}) and integrate with the \texttt{vblk\_sequence}. 

By binding the \texttt{free\_areas} field with the \texttt{struct page} metadata in Linux buddy system, we have implemented a Linux-style buddy system variant in seL4-based CapBuddy (referred to as CapBuddy-Linux Style). Based on the evaluation, we found that the Linux-style implementation has space for optimization when deallocation in batch (see \autoref{eval:microbenchmark}), which motivates us to implement an optimized, bitmap-based variant called CapBuddy-Standard. CapBuddy-Standard replaces the linked list in Linux Buddy for organizing metadata with bitmaps and a red-black tree. Such an approach imposes an overhead of tens of CPU cycles while effectively improving batching deallocation (see \autoref{eval:microbenchmark}). To further optimize CapBuddy-Standard, we implement ``CapBuddy-Stripped'', a streamlined variant of CapBuddy-Standard. CapBuddy-Stripped applies two optimizations. First, it organizes bitmaps by machine word through conditional compilation, avoiding repeated architecture-dependent checks during bitmap queries and updates. Second, it inlines frequently executed fastpath logic, including bitmap operations, buddy-merge checks, and updates to the largest available block. With a shorter execution path, CapBuddy-Stripped preserves scalability while reducing the performance gap with CapBuddy-Linux Style.

% We also implemented a stripped-down version of CapBuddy-Standard with aggressive optimizations, including fewer branches, less architecture-dependent logic (which affects bitmaps), and more inline functions. This variant, called ``CapBuddy-Stripped'', speeds up bitmap-based approaches by 50\% to make them comparable to CapBuddy-Linux Style while maintaining scalability.

\subsection{Back-end Daemon}

We use the threading utilities of seL4 LibOS to spawn the back-end daemon during CapBuddy's bootstrap. To apply the event-driven model, we bind an seL4 thread to an seL4 \emph{Notification}, which is a kernel object that receives asynchronous signals and wakes up the bound thread. The \emph{Notification} is also bound with a \emph{SchedContext}, which is a kernel object that allows the thread to execute on a specific CPU core for a period. When a signal from the front-end arrives, the \emph{Notification} directs the signal to the daemon thread, which then appends to the kernel scheduler for having obtained the \emph{SchedContext} of the \emph{Notification}. After execution, the \emph{SchedContext} will be taken from the thread until the next time it receives a signal, making it unable to execute proactively.

\subsection{Capability Bookkeeping Optimization.}
To optimize capability bookkeeping, we replaced the linear scan search in seL4 Buddy with a hierarchical bitmap-based search. Such a bitmap-based approach can speed up search operations using hardware instructions (e.g., ``count leading zeros''). seL4-based CapBuddy provides the capability bookkeeper with batching mechanisms, allowing the allocation of contiguous slots in CSpace for \emph{vblks}.

\subsection{Integrate CapBuddy into seL4 LibOS.}\label{impl:integrate}
\begin{figure}[htbp]
 \centering
  \includegraphics[width=0.6\columnwidth]{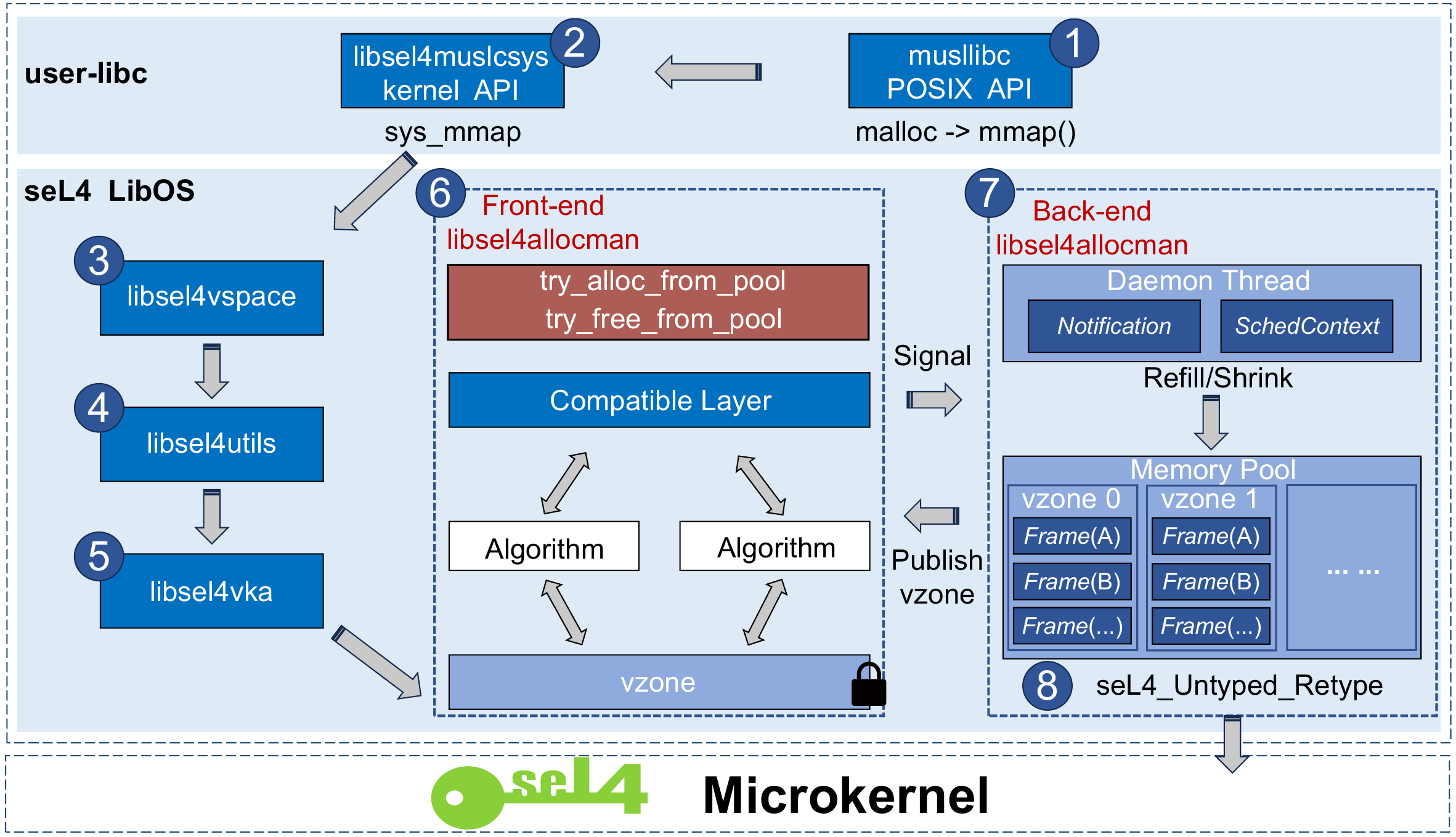}
 \caption{Implementation path of seL4-based CapBuddy.}
 \label{fig:sel4-buddy-retrofit}
\end{figure}

\autoref{fig:sel4-buddy-retrofit} shows how we prototype seL4-based CapBuddy on seL4 LibOS. In seL4 LibOS, client programs request memory through the POSIX APIs provided by \texttt{musllibc}. These requests pass through multiple LibOS layers before reaching \texttt{libsel4vka}, which is a library that provides a set of generic kernel-object allocation interfaces. \texttt{libsel4vka} forwards concrete allocation and deallocation requests to an allocator instance (i.e., \texttt{libsel4allocman} in our case) through function hooks.

To prototype CapBuddy into this framework, we add two hooks to \texttt{libsel4vka} that intercept requests for mappable memory from \texttt{libsel4allocman}: (1) \texttt{try\_alloc\_from\_pool\allowbreak ()}, which handles allocation requests for \emph{Frames}. (2) \texttt{try\allowbreak \_free\_from\_pool()}, which handles deallocation requests. These two hooks are the starting point to implement the front-end of CapBuddy, which serves requests with materialized \emph{vblks}.

We implement core functions of CapBuddy (e.g., buddy algorithm, \emph{vzone} management) as a library and integrate both the front-end and the back-end into \texttt{libsel4allocman}, which allows CapBuddy to reuse a majority of utilities for managing CSpace and \emph{Untyped} memory. With this design, \emph{Frame} requests issued through \texttt{libsel4vka} will first enter CapBuddy and go through the front-end, if they complete after a fastpath walk through the front-end returns to \texttt{libsel4vka}, otherwise the front-end will wait for the back-end daemon to refill the pool.

% To implement the front-end, the back-end, and the \emph{vzone} pool, we revamp \texttt{libsel4allocman} by splitting 

% Inside \texttt{libsel4allocman}, we split the original allocation stack into a front-end and a back-end. The front-end handles requests forwarded by the new hooks, while the back-end performs the \emph{vzone} provisioning. 
We know that the front-end signals the daemon when a threshold is met. When the front-end requires new provisioned \emph{vblks}, i.e., new \emph{Frame} capabilities, the back-end daemon receives signals and is scheduled to replenish the \emph{vzone} pool through \texttt{seL4\_Untyped\_Retype} syscall, which creates \emph{Frames} in batch for \emph{vzones}. This organization inserts CapBuddy into the existing seL4 LibOS memory management path while keeping the upper library interfaces unchanged.

\paragraph{\textbf{Large-Page Support.}}
We use the \emph{vzone} abstraction to support large pages in seL4-based CapBuddy, with separate \emph{vzones} and refill/shrink thresholds in the pool. In our prototype, each 32 MB large-page \emph{vzone} contains 16 \emph{vblks}, each corresponding to a 2 MB \emph{Frame}. The back-end prepares resources for these \emph{vzones}. Similar to Linux HugeTLB's management of free huge pages~\cite{linux_hugetlb}, the front-end maintains a linked list of free \emph{vblk} indices in each \emph{vzone}. Allocation removes an index from the list and returns the corresponding \emph{cslot} reference. Deallocation returns the index to the list. The interfaces in \autoref{impl:integrate} forward requests to new large-page allocation and deallocation hooks in \texttt{libsel4vka} based on the \emph{Frame} type and page size. During deallocation, we check whether the \emph{Frame} belongs to a large-page \emph{vzone}.

\section{Evaluation}\label{eval}

Given that we prototype CapBuddy on top of seL4 Buddy and seL4 LibOS, it is difficult to conduct a fair comparison against other OSes that are neither LibOS (i.e., Linux) nor managing physical memory by themselves (i.e., Unikraft). In this context, our evaluation focuses on the performance improvement CapBuddy brings to physical memory management in seL4 LibOS. We also use buddy systems in other OSes as performance references to assess the performance CapBuddy can achieve under capability-based resource management delegation.
Our evaluation answers the following questions.

\begin{itemize}
    \item How does CapBuddy improve physical memory allocation and deallocation performance?
    \item How does CapBuddy improve the end-to-end performance of user-level memory operations?
    \item How do emerging scenarios benefit from CapBuddy?
    \item How do real-world applications benefit from CapBuddy?
\end{itemize}

\paragraph{\textbf{Experimental Setup.}} We run all benchmarks on a Raspberry Pi 3B+ (64-bit) platform. This platform features a quad-core Cortex-A53 processor (fixed at 1.4 GHz) and 1 GB of DRAM. We choose it because seL4 is mature in the ecosystem of ARM processors~\cite{seL4_hw_support}. User space relies on musl libc to provide compatibility.

\paragraph{\textbf{Baselines.}} We run three seL4-based CapBuddy variants: CapBuddy-Standard, CapBuddy-Stripped, and CapBuddy-Linux Style. For baselines, we compare CapBuddy with:
\begin{itemize}
    \item \textbf{Linux Buddy (kernel version 6.12.48).} Represents physical memory management in a mature operating system. To minimize the impact of services on memory layout, it runs a minimal BusyBox (version 1.36.1).
    \item \textbf{seL4 Buddy.} The native allocator in seL4 LibOS. It does not support allocating or deallocating multiple physically contiguous 4 KB pages in a single request. We use it to quantify the performance improvement CapBuddy brings to the LibOS.
    \item \textbf{Unikraft Buddy.} Implements a minimal, pure buddy system algorithm that operates within a pre-established virtual address space. Compared to other physical memory managers, it is closer to the definition of a heap manager~\cite{unikraft21}. 
    Since it provides ideal performance but does not perform actual physical memory management, we consider it a theoretical optimum.
\end{itemize} 

%\subsection{Physical Memory Allocation Performance of the CapBuddy Back-end}
%We measure the latency of CapBuddy's back-end compared with seL4 Buddy when requesting physical memory from the kernel. The experiment allocated a total of 912 MB in 4 MB iterations, where each iteration retyped a single \emph{Untyped} into 1,024 4 KB pages until the total was reached. For each 4~MB allocation, the seL4 Buddy retyping by invoking \texttt{seL4\_Untyped\_Retype} individually for each page. In contrast, CapBuddy's back-end utilizes the batch argument within a single \texttt{seL4\_Untyped\_Retype} invocation to allocate the equivalent number of pages. The results show that CapBuddy takes an average of 51.5~$\mu$s to allocate 4~MB, whereas seL4 Buddy takes 1,052.8~$\mu$s. Using this batching mechanism, the CapBuddy's back-end achieves 20.5 times speedup in requesting physical memory compared to seL4 Buddy.

\subsection{Microbenchmarks for Physical Memory Allocation and Deallocation}\label{eval:microbenchmark}
We evaluate each buddy system for both single-page and contiguous physical memory allocation and deallocation. The aim is to demonstrate the latency of a buddy system in responding to a memory request from other OS services. We also compare CapBuddy with seL4 Buddy for large-page allocation and deallocation.

\begin{figure*}[!t]
    \centering
    \includegraphics[width=\textwidth]{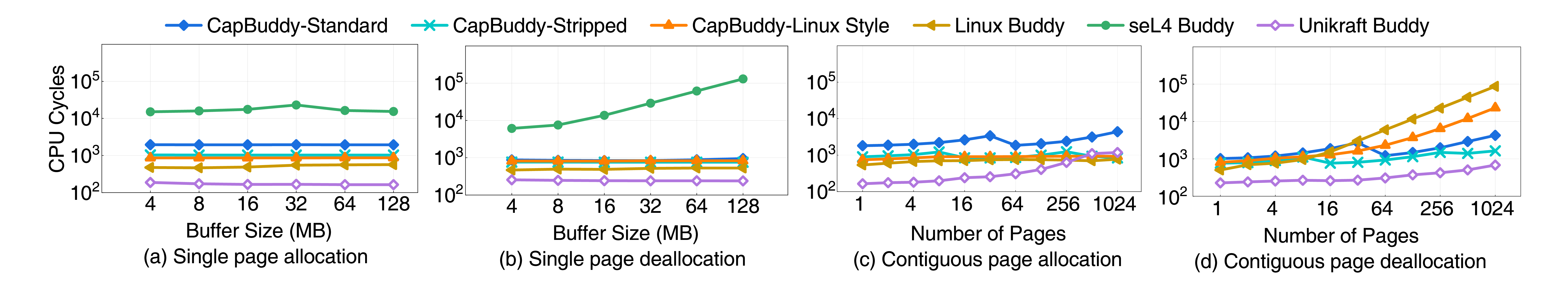}
    \caption{
    (a)–(b) show the results for single-page physical memory allocation and deallocation; (c)–(d) show the results for contiguous physical memory allocation and deallocation.
    }    
    \label{fig:mirco}
\end{figure*}

\paragraph{\textbf{Single-Page Evaluation.}} We measure the latency of single-page allocation and deallocation by allocating and deallocating 4 KB pages (a total amount ranging from 4 MB to 128 MB). \autoref{fig:mirco} (a) and \autoref{fig:mirco} (b) show the results. Overall, Unikraft Buddy exhibits the lowest latency, followed by Linux Buddy. seL4 Buddy performs the worst, while CapBuddy falls between Linux Buddy and seL4 Buddy.

%Figure (a) and Figure (b) show the results for single-page physical memory allocation and deallocation, while Figure (c) and Figure (d) show the results for contiguous physical memory allocation and deallocation. Lower is better in all figures.

The single-page allocation results (\autoref{fig:mirco} (a)) show that all three CapBuddy variants achieve significant performance improvements over seL4 Buddy. seL4 Buddy's allocation latency reaches 15,111 $\sim$ 23,176 cycles. In contrast, CapBuddy-Standard, CapBuddy-Stripped, and CapBuddy-Linux Style achieve average performance improvements of 8.9 times, 16.8 times, and 20.0 times over seL4 Buddy, respectively.

Among all CapBuddy variants, CapBuddy-Linux Style (860 $\sim$ 870 cycles) performs best, though it is still 40.1\% slower than native Linux. This remaining gap mainly comes from the seL4 LibOS compatibility layer used by our prototype, which typically takes 330 cycles. However, when comparing to Unikraft, i.e., the theoretical optimum (164 $\sim$ 188 cycles), both CapBuddy-Linux Style and native Linux are way slower. At their lowest mean allocation latencies of 1,941 and 860 cycles, CapBuddy-Standard and CapBuddy-Linux Style achieve approximately $0.7$ and $1.6$ million pages/s, respectively.

The single-page deallocation results (\autoref{fig:mirco} (b)) follow a similar trend to allocation. seL4 Buddy's latency is 6,119 $\sim$ 130,339 cycles, which is significantly higher than that of other systems. CapBuddy-Standard, CapBuddy-Stripped, and CapBuddy-Linux Style achieve average performance improvements of 45.6 times, 54.5 times, and 50.0 times over seL4 Buddy, respectively. Here, CapBuddy-Stripped performs best (760 $\sim$ 765 cycles), slower than Linux Buddy's average performance (469 $\sim$ 530 cycles) and Unikraft Buddy's (239 $\sim$ 255 cycles).

\paragraph{\textbf{Batching Evaluation.}} We measure allocation and deallocation latency for block sizes ranging from 4 KB to 4 MB. For each size, we allocate a total of 256 MB. We exclude the seL4 Buddy from this evaluation as it does not support batching.

\autoref{fig:mirco} (c) shows the results for batching allocation. CapBuddy-Linux Style outperforms other CapBuddy variants with latencies of 744 $\sim$ 936 cycles. It achieves 78.8\% of the Linux Buddy's average performance (538 $\sim$ 774 cycles) and 48.5\% of the Unikraft Buddy's average performance (164 $\sim$ 1,183 cycles). CapBuddy-Stripped and CapBuddy-Standard exhibit latencies of 838 $\sim$ 1,234 cycles and 1,818 $\sim$ 4,399 cycles, respectively.

\autoref{fig:mirco} (d) shows the results for batching deallocation. CapBuddy outperforms Linux Buddy in this metric. CapBuddy-Stripped (765 $\sim$ 1,641 cycles) performs best among other CapBuddy variants. It outperforms the Linux Buddy by 11.5 times on average but achieves only 32.0\% of the Unikraft Buddy's performance (228 $\sim$ 684 cycles). CapBuddy-Linux Style exhibits high latency variance (808 $\sim$ 23,189 cycles), whereas CapBuddy-Standard maintains latencies between 1,016 and 4,252 cycles.

\begin{figure}[htbp]
    \centering
    \includegraphics[width=0.5\textwidth]{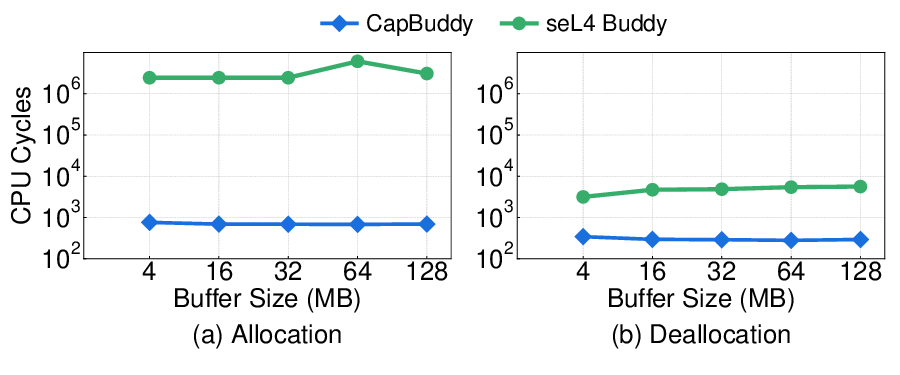}
    \caption{(a) shows the results for large-page physical memory allocation; (b) shows the results for large-page physical memory deallocation.}
    \label{fig:large_page}
\end{figure}

\paragraph{\textbf{Large-Page Evaluation.}} We further measure allocation and deallocation latency for 2 MB pages (a total amount ranging from 4 MB to 128 MB). \autoref{fig:large_page} (a) and \autoref{fig:large_page} (b) show the results. CapBuddy achieves allocation latencies of 680 $\sim$ 761 cycles, whereas seL4 Buddy requires 2.44M $\sim$ 6.11M cycles. Deallocation follows a similar trend, with CapBuddy achieving latencies of 279 $\sim$ 343 cycles, lower than seL4 Buddy's 3,155 $\sim$ 5,613 cycles.

\paragraph{\textbf{Overall Analysis.}} In the single-page and batching evaluations, Unikraft Buddy achieves the best performance because it has a minimal code path, as it operates purely as a heap allocator, i.e., not managing physical memory. 

In batching deallocation, CapBuddy (Standard/Stripped) outperforms Linux. The reason Linux underperforms is that it iterates over per-page metadata and performs additional checks when freeing contiguous blocks. CapBuddy-Linux Style still retains this property, leading to higher latency. In contrast, the bitmap-based CapBuddy variants (Standard and Stripped) eliminate metadata traversal, resulting in lower latency.

Although most of the CapBuddy variants underperform Linux in single-page evaluations by 300 $\sim$ 400 cycles, such an overhead is solvable with a slimmed compatibility layer.

\subsection{Back-end Daemon Overhead}\label{eval:daemon-overhead}

Using CapBuddy-Standard, we measure back-end \emph{vzone} refill and shrink overhead, synchronous fallback latency when the \emph{vzone} pool has insufficient memory, and the daemon's scheduling interference with a task on the same core.

\paragraph{\textbf{Vzone Refill and Shrink.}} We refill and shrink a total of 512 MB in units of 4 MB \emph{vzones}, measuring 128 operations each. We measure overall and stage latencies together to identify the main sources of overhead.

\autoref{tab:daemon-overhead} shows the overall latency and stage breakdown. A few operations spend much longer in the \emph{Untyped} acquisition and contiguous \emph{cslot} allocation stage, raising mean refill latency far above the median, while \emph{Frame} creation dominates most refill operations. Shrink latency is relatively stable, dominated by \emph{Frame} capability and \emph{cslot} reclamation. CapBuddy offloads these operations to the back-end, so the front-end allocates and deallocates by updating \emph{vblk} states without the corresponding capability operations on the fastpath.

\begin{table}[htbp]
    \centering
    \small
    \renewcommand{\arraystretch}{1.15}
    \setlength{\tabcolsep}{5pt}
    {
    \captionsetup{skip=4pt}
    \caption{Latency of \emph{vzone} refill, shrink, and synchronous fallback (in CPU cycles; k = $10^3$, M = $10^6$).}
    \label{tab:daemon-overhead}
    \begin{tabular}{@{}lrrr@{}}
        \toprule[1pt]
        Operation / Stage & Mean & Median & Max \\
        \midrule
        \textbf{Vzone refill (overall)}                          & 6.25M & 91.8k & 315M \\
        \quad \emph{Untyped} acquisition and contiguous \emph{cslot} allocation
                                                                & 6.17M & 8.60k & -- \\
        \quad \emph{Frame} creation                              & 72.5k & 73.1k & -- \\
        \quad \emph{vzone} initialization and publication         & 12.1k & 9.74k & -- \\
        \quad Other                                             & 310   & 296   & -- \\
        \midrule
        \textbf{Vzone shrink (overall)}                          & 445k  & 444k  & 462k \\
        \quad \emph{vzone} detachment                            & 2.20k & 2.24k & -- \\
        \quad \emph{Frame} capability and \emph{cslot} reclamation & 434k  & 434k  & -- \\
        \quad \emph{Untyped} return and metadata release          & 7.30k & 6.84k & -- \\
        \quad Other                                             & 1.09k & 1.04k & -- \\
        \midrule
        \textbf{Synchronous fallback (overall)}                  & 6.26M & 100k  & 315M \\
        \bottomrule[1pt]
    \end{tabular}
    }
\end{table}

\paragraph{\textbf{Synchronous Fallback Path.}} Starting with an empty \emph{vzone} pool, we continuously allocate 4 KB pages totaling 512 MB. Each time the pool is exhausted, we measure synchronous fallback latency from the front-end receiving the request to returning the page after synchronous refill. The 128 synchronous fallback operations have a mean latency close to that of refill (\autoref{tab:daemon-overhead}). When the pool cannot satisfy a request, the front-end waits for refill, bringing the back-end's resource preparation onto the request's critical path.

\paragraph{\textbf{CPU Scheduling Interference.}} We bind CoreMark~\cite{eembc_coremark}, a compute-intensive benchmark, and the high-priority daemon to CPU 1, with the memory allocation client on CPU 0. To measure interference on the same core, we compare CoreMark throughput with the daemon idle or performing back-end operations. Each measurement lasts 20 seconds. With back-end operations, the pool initially contains one 4 MB \emph{vzone}. After measurement starts, the client allocates 512 MB in 4 KB page requests, then deallocates the pages; the daemon reclaims the \emph{vzones}. This triggers multiple \emph{vzone} refill and shrink operations, reducing CoreMark throughput by 3.1\% (\autoref{tab:daemon-interference}).

\begin{table}[htbp]
    \centering
    \small
    \renewcommand{\arraystretch}{1}
    \sisetup{group-separator={,},group-minimum-digits=4}
    {
    \captionsetup{skip=4pt}
    \caption{Daemon interference with CoreMark throughput on the same core. We report the median throughput of 10 runs for each condition.}
    \label{tab:daemon-interference}
    \begin{tabular}{lS[table-format=4.1]}
        \toprule[1pt]
        Condition & {\shortstack{CoreMark Throughput (iterations/s)}} \\
        \midrule
        Idle daemon                                    & 4593.4 \\
        Daemon performing refill and shrink            & 4452.0 \\
        \bottomrule[1pt]
    \end{tabular}
    }
\end{table}

\begin{figure}[htbp]
    \centering
    \includegraphics[width=0.95\textwidth]{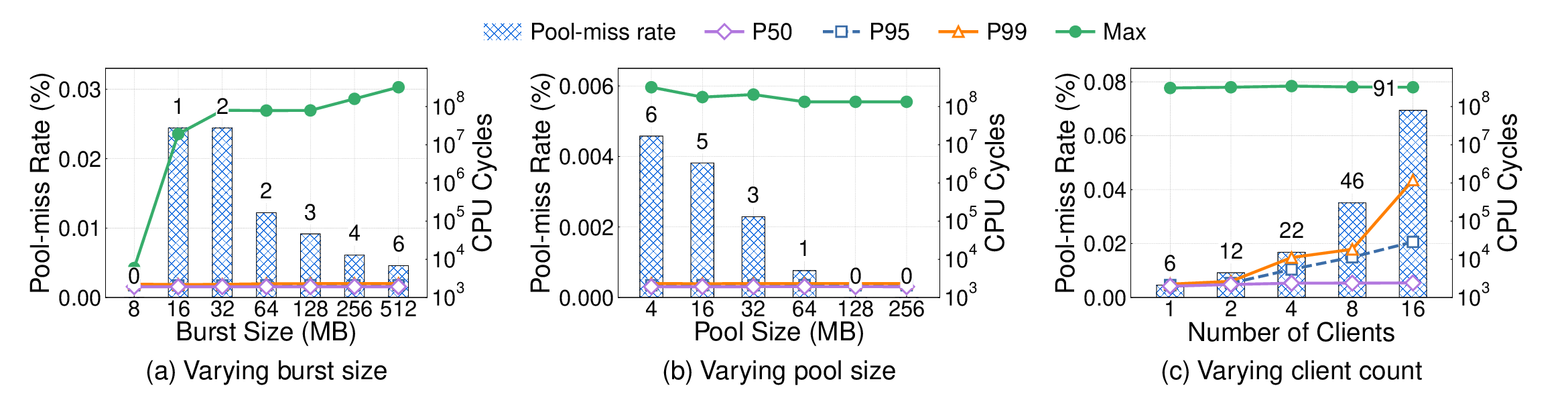}
    \caption{Results for different (a) burst sizes, (b) initial pool sizes, and (c) client counts. Bars show the pool-miss rate (left axis), with pool-miss counts above. Lines show P50, P95, P99, and Max allocation latencies in CPU cycles (right axis).}
    \label{fig:burst_evaluation}
\end{figure}

\subsection{Burst Requests}
We test whether asynchronous refill in CapBuddy-Standard satisfies single-client and multi-client burst requests in time with insufficient initial \emph{vzone} pool memory. We record P50, P95, P99, and maximum (Max) allocation latencies and pool-miss rate: the percentage of allocation requests unable to obtain a page from the pool (pool misses).

\paragraph{\textbf{Single-Client Burst Requests.}} We bind the client to CPU 0 and the daemon to CPU 1. The client continuously allocates 4 KB pages; the burst size is the total memory requested per batch. We vary the burst size from 8 MB to 512 MB with an initial 4 MB \emph{vzone} pool, then vary the initial pool size from 4 MB to 256 MB with a fixed 512 MB burst.

In both experiments (\autoref{fig:burst_evaluation} (a) and (b)), P50 remains at 1,913 $\sim$ 1,916 cycles, P95 at 2,067 $\sim$ 2,098 cycles, and P99 at 2,175 $\sim$ 2,324 cycles. With an initial 4 MB pool, the 8 MB and 512 MB bursts cause 0 and 6 pool misses, respectively; all pool-miss rates remain below 0.025\%. For a fixed 512 MB burst, larger initial pools reduce pool misses, with none at 128 MB and 256 MB but still high Max. Asynchronous refill thus satisfies most front-end requests in time. Increasing the burst size does not noticeably affect median allocation latency. Max remains high without pool misses because interrupts cause the front-end core to enter seL4's SMP kernel, which uses a big kernel lock (BKL) to protect shared kernel state~\cite{Peters_DEH_15}. If the kernel on the daemon's core holds the BKL, the front-end core must wait, delaying allocation.

\paragraph{\textbf{Multi-Client Burst Requests.}} We extend the single-client prototype to multiple client threads sharing one \emph{vzone} pool and one refill daemon, with pool-level and \emph{vzone}-level locks protecting concurrent accesses. Clients send no additional refill notifications while a refill request is pending or being processed. On CPUs 0, 2, and 3, 1 to 16 clients concurrently request a total of 512 MB in 4 KB pages from an initial 4 MB pool; the daemon runs on CPU 1.

From 1 to 16 clients (\autoref{fig:burst_evaluation} (c)), pool misses increase from 6 to 91, and pool-miss rate rises from 0.0046\% to 0.0694\%. P50 allocation latency increases from 1,973 to 2,407 cycles, P95 from 2,100 to 28,535 cycles, and P99 from 2,236 to 1.19M cycles. Most requests complete quickly, but more clients cause more frequent pool misses and higher tail latency.

\subsection{Microbenchmarks for User-Level End-to-End Memory Operation}
This section evaluates the end-to-end latency of memory operations from the user-level perspective. In applications, a complete memory request involves the allocation/deallocation of physical memory and the corresponding mapping/unmapping operations. To measure the above requests, we conduct three microbenchmarks: ``single-page'', ``batching'', and a memory bandwidth evaluation. In the first two benchmarks, Linux uses \texttt{mmap} with \texttt{MAP\_POPULATE} (for immediate allocation and mapping) and \texttt{munmap}, while the seL4 LibOS uses the semantically equivalent \texttt{vspace\_new\_pages} and \texttt{vspace\_unmap\_pages}. We exclude Unikraft for these experiments because we did not find similar Unikraft mechanisms to compare against.

\paragraph{\textbf{Single-Page Evaluation.}} We measure the end-to-end latency for single-page (4 KB) operations along the memory management path. \autoref{tab:mmap} shows that CapBuddy-Standard has the best performance: it outperforms seL4 Buddy by 4.8 times and Linux Buddy by 2.6 times in terms of ``allocation + mapping'', while for ``deallocation + unmapping'', CapBuddy-Standard outperforms seL4 Buddy by 1.9 times and Linux Buddy by 2.5 times.

%Overall, CapBuddy-Standard demonstrates superior performance in the single-page evaluation. 
Compared with seL4 Buddy, the reduced latency stems primarily from faster single-page allocation and deallocation. Compared to Linux Buddy, the performance gain comes from the shorter and more compact virtual memory management path of the seL4 LibOS.

\begin{table}[htbp]
\centering
    \small
    \renewcommand{\arraystretch}{1}
    \begin{tabular}{lS[table-format=5]S[table-format=5]}
        \toprule[1pt]
        Evaluated System  & \text{Alloc + Mmap} & \text{Dealloc + Unmap} \\ 
        \midrule
        Linux Buddy            & 14810    & 14389 \\  
        CapBuddy-Standard      & 5716     & 5731 \\ 
        seL4 Buddy             & 27689    & 11076 \\ 
        \bottomrule[1pt]
    \end{tabular}
    \captionsetup{skip=4pt}
    \caption{Latency of single-page end-to-end memory operations (in CPU cycles).}
    \label{tab:mmap}
\end{table}

\begin{figure}[htbp]
    \centering
    \includegraphics[width=\textwidth]{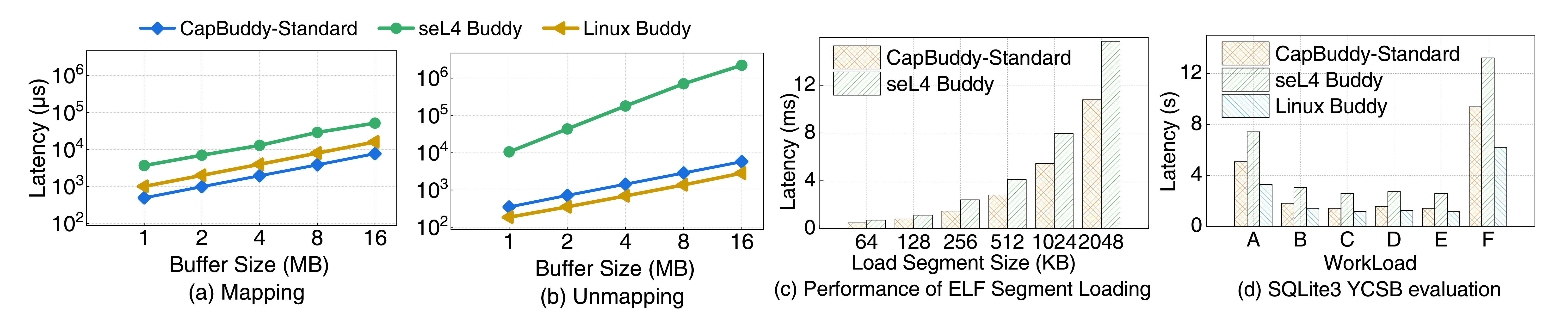}
    \caption{(a) shows the results for batching ``allocation + mapping''; (b) shows the results for batching ``unmapping + deallocation''; (c) shows the results of the emerging scenario simulation; (d) shows the results of SQLite3 with YCSB workload.}
    \label{fig:mmap_elf_and_sqlite}
\end{figure}

%Figure (a) shows the results for batching "allocation + mapping", and Figure (b) shows the results for batching "deallocation + unmapping". Lower is better in both figures.

\paragraph{\textbf{Batching Evaluation.}} We measure end-to-end latency of batching ``allocation/deallocation + mapping/unmapping'' of varying sizes, ranging from 1 MB to 16 MB in powers of two. \autoref{fig:mmap_elf_and_sqlite} (a) and \autoref{fig:mmap_elf_and_sqlite} (b) show that CapBuddy-Standard achieves the best performance across most operations. 

For ``allocation + mapping'' (i.e., \autoref{fig:mmap_elf_and_sqlite} (a)), CapBuddy-Standard exhibits latencies ranging from 490  $\sim$  7,687~\textmu s. On average, it outperforms seL4 Buddy (3,678 $\sim$ 51,382~\textmu s) by 7.1 times and Linux Buddy (1,003 $\sim$ 15,918~\textmu s) by 2.1 times.

For ``deallocation + unmapping'' (i.e., \autoref{fig:mmap_elf_and_sqlite} (b)), CapBuddy-Standard latencies range from 350  $\sim$  5,743~\textmu s. It outperforms seL4 Buddy by 169.9 times on average, but achieves only 49.7\% of Linux Buddy's performance.

CapBuddy-Standard outperforms seL4 Buddy in this benchmark primarily because it supports contiguous memory allocation/deallocation. For allocation + mapping, CapBuddy-Standard surpasses Linux due to its shorter virtual memory management path. However, CapBuddy-Standard underperforms Linux during deallocation + unmapping, which is attributed to mature micro-architectural optimizations in the Linux kernel (e.g., TLB batching), which provide significant advantages in batch reclamation scenarios.

\begin{figure*}[!t]
    \centering
    \includegraphics[width=\textwidth]{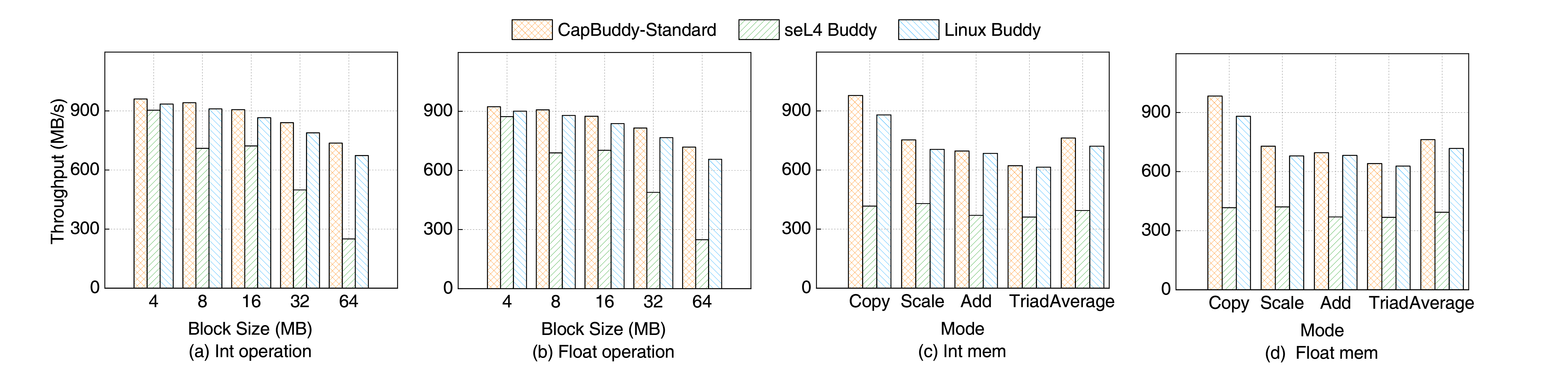}
    \caption{Results of RAMspeed.}
    \label{fig:ramspeed}
\end{figure*}

%Figures (a)–(d) show the RAMspeed benchmark results, higher is better; Figure (e) shows the results of the emerging scenario simulation, lower is better; Figure (f) shows the results of SQLite3 with YCSB workload, lower is better.

\paragraph{\textbf{Bandwidth Evaluation.}} RAMspeed is a benchmark that measures the bandwidth of memory subsystems~\cite{ramspeed}. We use it to evaluate how different memory managements affect the measured bandwidth. We use six ARM workloads provided by RAMspeed, allocating a total of 256 MB of memory. The motivation is that it helps demonstrate how memory management performs in responding to requests for contiguous memory. The RAMspeed benchmark allocates large, power-of-two chunks, which trigger anonymous mapping and unmapping in musl libc. These operations correspond to the allocation, mapping, deallocation, and unmapping of contiguous physical memory.

\autoref{fig:ramspeed} shows that CapBuddy-Standard achieves the best performance. Compared to the seL4 Buddy, CapBuddy-Standard improves overall RAMspeed performance by 77.0\% on average, with a maximum speedup of 1.9 times. Compared with Linux Buddy, it improves average performance by 5.4\%, with a maximum improvement of 11.6\%. This advantage primarily stems from the batching mechanism of CapBuddy. 

% \subsection{The Demands of Short-Lived Applications in Emerging Scenarios}
\subsection{ELF Loading}
Previous evaluations can help answer how memory management may affect runtime performance (e.g., alloc + mapping), which is critical for long-lived applications. However, a key characteristic of emerging scenarios is the need for thread spawning, which is critical for the performance of short-lived applications (e.g., serverless~\cite {Szekely_BMK_24}). Such requirements rely on loading executables on the fly to assemble execution contexts. For the above reasons, we measure the performance of ELF loading, which is crucial for thread spawning, to show how emerging scenarios can benefit from CapBuddy. 

For ELF loading, we evaluate segment sizes ranging from 64 KB to 2 MB, comparing the performance of CapBuddy-Standard with seL4 Buddy. \autoref{fig:mmap_elf_and_sqlite} (c) shows that CapBuddy improves performance by an average of 32.3\% over seL4 Buddy, with a maximum improvement of 39.1\%. %For thread spawning,  %These results demonstrate that CapBuddy-Standard enhances seL4 LibOS performance in scenarios requiring multi-process support.

\subsection{Real-world Application}
We port SQLite3~\cite{SQLite} to both Linux and seL4 LibOS, along with the log-structured file system (xv6fs~\cite{xv6fs}) and a RAM Drive acting as a block device to support SQLite3. We use SQLite3 to measure CapBuddy's impact on database operations. The evaluation over SQLite3 in its default configuration (logging enabled) using the six workloads (A through F) defined by YCSB~\cite{ycsb}. Each workload comprises a different mixture of read/write/update/scan operations. To minimize disk I/O, we enabled SQLite3's in-memory database feature.

We never claim that SQLite3 is optimal, as it relies on the entire OS to deliver good performance. For this reason, we use SQLite3 to assess how well CapBuddy improves overall performance through a memory management boost.

\autoref{fig:mmap_elf_and_sqlite} (d) shows that CapBuddy-Standard achieves an average performance improvement of 38.9\% over seL4 Buddy, with a maximum improvement of 44.9\%. Compared to Linux, CapBuddy-Standard achieves an average of 75.4\% of its performance, with a maximum of 81.0\%. Regarding the performance gap with Linux, we observe that SQLite3 frequently triggers \texttt{sbrk} to expand the heap during execution. Linux uses demand paging for \texttt{sbrk}, whereas seL4 LibOS relies on proactive mapping, which explains the performance gap.

% \begin{figure}[H]
%     \centering
%     \includegraphics[width=0.45\textwidth]{Picture/load_sqlite.eps}
%     \caption{Figure (a) shows the results of the emerging scenario simulation, and Figure (b) shows the performance of SQLite3 under YCSB workloads.}
%     \label{fig:load_sqlite}
% \end{figure}

% \clearpage
% \begin{figure}[H]   
%     \centering
%     \includegraphics[width=\textwidth]{Picture/Fig10.eps}
%     %\caption{Basic Continuous Evaluation with Nobacked}
%     \label{fig:Fig10}
% \end{figure}

% \begin{figure}[H]   
%     \centering
%     \includegraphics[width=0.5\textwidth]{Picture/Fig11.eps}
%     %\caption{Basic Continuous Evaluation with Nobacked}
%     \label{fig:Fig11}
% \end{figure}

%  \centering
%     \includegraphics[width=\textwidth]{Picture/Fig1.1.eps}
%     \caption{Basic Continuous Evaluation with Nobacked}
%     \label{fig:Fig1}
% \end{figure}

% \begin{figure}[H]   
%     \centering
%     \includegraphics[width=0.5\textwidth]{Picture/Fig13.eps}
%     %\caption{Basic Continuous Evaluation with Nobacked}
%     \label{fig:Fig13}
% \end{figure}

%  \begin{figure}[H]   
%     \centering
%     \includegraphics[width=0.5\textwidth]{Picture/Fig14.eps}
%     %\caption{Basic Continuous Evaluation with Nobacked}
%     \label{fig:Fig14}
% \end{figure}

\section{Related Work}

Most of the members of the L4 family employ recursive page mappings for physical memory management~\cite{ukernel95}, which bind physical memory to page mappings. Though effective, this approach introduces an in-kernel mapping database, which consumes a large amount of memory (25\% $\sim$ 50\% of kernel heap). The drawback includes allowing malicious processes to create mappings to DoS attack the kernel heap. %Liedtke proposed the clans-and-chiefs method \cite{liedtke1992clans}. 
Nowadays, some microkernels (e.g., L4Re~\cite{Hartig_HLSW_97}) choose to stay with recursive page mappings, while seL4  abandoned this approach by replacing it with an Exokernel-like design, i.e., providing capability-protected kernel abstractions that adhere closely to the property of physical memory. 

In terms of the user-level memory management over such capability-based resource-controlling microkernels, Elkaduwe et al.~\cite{sel4mm-phd10} evaluated slab, buddy, and hybrid allocators on seL4, finding that buddy allocation incurs high memory overhead (up to 200\%) for small kernel objects and recommending slab allocation. Eftime et al.~\cite{eftime2017optimizing} proposed a bitmap-based allocator to reduce metadata and accelerate small memory allocations using bitwise operations.

% In terms of physical memory management, Elkaduwe evaluated multiple memory allocation methods on seL4 \cite{elkaduwe2010principled}, including slab allocation, buddy allocation, and hybrid allocation (running a slab allocator on a buddy system). The evaluation results show that the buddy allocator causes an extremely high memory load for allocating small kernel objects due to its use of many capabilities, even reaching 200\%, so using a slab allocator for seL4 is recommended. Eftime et al. \cite{eftime2017optimizing} designed a bitmap-based allocator to reduce the metadata used by memory management and improve the allocation performance of small memory blocks. They reduced metadata and utilized bit operation instructions provided by modern architecture to accelerate.

The buddy system~\cite{knowlton1965fast}, widely adopted for its high performance and low external fragmentation, has inspired several variants, such as Fibonacci buddies~\cite{hirschberg1973class}, generalized Fibonacci buddies~\cite{burton1976buddy, peterson1977buddy}, weighted buddies~\cite{shen1974weighted}, and double buddies~\cite{page1986improving, wilson1995memory}. Recent research has focused on enhancing performance on multi-core platforms. The lazy iBuddy system~\cite{park2014ibuddy} optimized single-page allocation and reduced lock contention by using individual page frames and multiple buddy space management. NBBS~\cite{marotta2021nbbs} enabled non-blocking concurrent allocations and releases, avoiding spin-lock synchronization. Spring Buddy~\cite{lu2021spring} tackled memory allocation bottlenecks in cloud computing by dynamically adjusting page frame organization based on request patterns.
\section{Conclusion}
Capability-based resource-controlling microkernels protect hardware resources with a capability system to securely expose low-level interfaces to the user space, aimed at improving application performance without sacrificing security. However, we find that the cost of the capability system can impose overhead on user-level dynamic memory management. The bottleneck is the long critical path of memory request handling. We propose CapBuddy, a first step towards a dynamic memory management approach for a capability-based resource-controlling microkernel, which offloads time-consuming capability-relevant operations from the critical path. A prototype of CapBuddy on seL4 LibOS improves the performance of dynamic memory management by 20 to 54.5 times, bringing it to the same order of magnitude as the Linux buddy system.

\begin{acks}
This work is supported by the National Natural Science Foundation of China (No.62272458 and No.62372311), and the National Natural Science Foundation of China for Distinguished Young Scholars (No.62125208).
\end{acks}

\bibliographystyle{ACM-Reference-Format}
\bibliography{reference}

@inproceedings{mmimpact93,
  title={The impact of operating system structure on memory system performance},
  author={Chen, J Bradley and Bershad, Brian N},
  booktitle={Proceedings of the 14th ACM Symposium on Operating System Principles (SOSP)},
  pages={120--133},
  publisher={{ACM}},
  doi={10.1145/168619.168629},
  year={1993}
}

@inproceedings{sel4ten13,
  title={From {L3} to {seL4} what have we learnt in 20 years of {L4} microkernels?},
  author={Elphinstone, Kevin and Heiser, Gernot},
  booktitle={Proceedings of the 24th ACM Symposium on Operating Systems Principles (SOSP)},
  pages={133--150},
  year={2013},
  publisher={{ACM}},
  doi={10.1145/2517349.2522720}
}

@inproceedings{Lyons_MAH_18,
    author    = {Lyons, Anna and McLeod, Kent and Almatary, Hesham and Heiser, Gernot},
    title     = {Scheduling-Context Capabilities: A Principled, Light-Weight
                 Operating-System Mechanism for Managing Time},
    booktitle = {Proceedings of the Thirteenth EuroSys Conference},
    series    = {EuroSys '18},
    year      = {2018},
    pages     = {1--16},
    publisher = {Association for Computing Machinery},
    address   = {New York, NY, USA},
    location  = {Porto, Portugal},
    doi       = {10.1145/3190508.3190539}
}

@inproceedings{ukernel95,
  title={On micro-kernel construction},
  author={Liedtke, Jochen},
  booktitle={Proceedings of the 15th ACM Symposium on Operating Systems Principles (SOSP)},
  pages={237--250},
  doi={10.1145/224056.224075},
  publisher={{ACM}},
  year={1995}
}

@inproceedings{Baumann19,
author = {Baumann, Andrew and Appavoo, Jonathan and Krieger, Orran and Roscoe, Timothy},
title = {A fork() in the road},
year = {2019},
isbn = {9781450367271},
publisher = {Association for Computing Machinery},
address = {New York, NY, USA},
url = {https://doi.org/10.1145/3317550.3321435},
doi = {10.1145/3317550.3321435},
booktitle = {Proceedings of the Workshop on Hot Topics in Operating Systems},
pages = {14–22},
numpages = {9},
location = {Bertinoro, Italy},
series = {HotOS '19}
}

@inproceedings{ukl23,
   title={Unikernel Linux (UKL)},
   url={http://dx.doi.org/10.1145/3552326.3587458},
   DOI={10.1145/3552326.3587458},
   booktitle={Proceedings of the Eighteenth European Conference on Computer Systems},
   publisher={ACM},
   author={Raza, Ali and Unger, Thomas and Boyd, Matthew and Munson, Eric B and Sohal, Parul and Drepper, Ulrich and Jones, Richard and De Oliveira, Daniel Bristot and Woodman, Larry and Mancuso, Renato and Appavoo, Jonathan and Krieger, Orran},
   year={2023},
   month=May,
   pages={590–605},
   collection={EuroSys ’23}
}

@inproceedings{vICirqCoalescing11,
    author = {Ahmad, Irfan and Gulati, Ajay and Mashtizadeh, Ali},
    title = {{vIC}: interrupt coalescing for virtual machine storage device {IO}},
    year = {2011},
    publisher = {USENIX Association},
    address = {USA},
    booktitle = {Proceedings of the 2011 USENIX Conference on USENIX Annual Technical Conference},
    pages = {4},
    numpages = {1},
    location = {Portland, OR},
    series = {USENIXATC'11}
}

@inproceedings{Ballooning02,
    author = {Waldspurger, Carl A.},
    title = {Memory resource management in VMware ESX server},
    year = {2002},
    isbn = {9781450301114},
    publisher = {USENIX Association},
    address = {USA},
    booktitle = {Proceedings of the 5th Symposium on Operating Systems Design and Implementation},
    pages = {181–194},
    numpages = {14},
    location = {Boston, Massachusetts},
    series = {OSDI '02}
}

@phdthesis{cherios21,
  title={{CheriOS}: designing an untrusted single-address-space capability operating system utilising capability hardware and a minimal hypervisor},
  author={Esswood, Lawrence},
  year={2021},
  school={University of Cambridge},
  address={United Kingdom}
}

@inproceedings{spliter25,
  title={Serverless functions made confidential and efficient with split containers},
  author={Shi, Jiacheng and Gu, Jinyu and Xia, Yubin and Chen, Haibo},
  booktitle={Proceedings of the 34th USENIX Security Symposium (USENIX Security)},
  pages={1091--1110},
  publisher={{USENIX} Association},
  year={2025}
}

@inproceedings{keykos92,
  title={The {KeyKOS} nanokernel architecture},
  author={Bomberger, Alan C and Hardy, Norman and Frantz, A Peri and Landau, Charles R and Frantz, William S and Shapiro, Jonathan S and Hardy, Ann C},
  booktitle={Proceedings of the USENIX Workshop on Micro-kernels and Other Kernel Architectures},
  pages={95--112},
  publisher={{USENIX} Association},
  year={1992}
}

@inproceedings{l4cap09,
  title={Taming subsystems: capabilities as universal resource access control in {L4}},
  author={Lackorzynski, Adam and Warg, Alexander},
  booktitle={Proceedings of the 2nd Workshop on Isolation and Integration in Embedded Systems (IIES)},
  pages={25--30},
  publisher={{ACM}},
  doi={10.1145/1519130.1519135},
  year={2009}
}

@inproceedings{eros99,
  title={{EROS}: a fast capability system},
  author={Shapiro, Jonathan S and Smith, Jonathan M and Farber, David J},
  booktitle={Proceedings of the 17th ACM Symposium on Operating Systems Principles (SOSP)},
  pages={170--185},
  publisher={{ACM}},
  doi={10.1145/319151.319163},
  year={1999}
}

@inproceedings{exokernel95,
  title={Exokernel: An operating system architecture for application-level resource management},
  author={Engler, Dawson R and Kaashoek, M Frans and O'Toole Jr, James},
  booktitle={Proceedings of the 15th ACM Symposium on Operating System Principles (SOSP)},
  pages={251--266},
  year={1995},
  doi={10.1145/224056.224076},
  publisher={{ACM}}
}

@inproceedings{exokernel97,
  title={Application performance and flexibility on exokernel systems},
  author={Kaashoek, M Frans and Engler, Dawson R and Ganger, Gregory R and Brice{\~{n}}o, H{\'e}ctor M and Hunt, Russell and Mazi{\`{e}}res, David and Pinckney, Thomas and Grimm, Robert and Jannotti, John and Mackenzie, Kenneth},
  booktitle={Proceedings of the 16th ACM Symposium on Operating System Principles (SOSP)},
  pages={52--65},
  publisher={{ACM}},
  doi={10.1145/268998.266644},
  year={1997}
}

@article{lionsos25,
  title={Fast, Secure, Adaptable: {LionsOS} Design, Implementation and Performance},
  author={Heiser, Gernot and Velickovic, Ivan and Chubb, Peter and Joshy, Alwin and Ganesh, Anuraag and Nguyen, Bill and Li, Cheng and Darville, Courtney and Zhu, Guangtao and Archer, James and Zhou, Jingyao and Winter, Krishnan and Parker, Lucy and Duchniewicz, Szymon and Bai, Tianyi},
  journal={CoRR},
  volume={abs/2501.06234},
  doi={10.48550/arxiv.2501.06234},
  eprinttype={arXiv},
  eprint={2501.06234},
  year={2025}
}

@inproceedings{hm24,
  title={Microkernel goes general: Performance and compatibility in the {HongMeng} production microkernel},
  author={Chen, Haibo and Miao, Xie and Jia, Ning and Wang, Nan and Li, Yu and Liu, Nian and Liu, Yutao and Wang, Fei and Huang, Qiang and Li, Kun and Yang, Hongyang and Wang, Hui and Yin, Jie and Peng, Yu and Xu, Fengwei},
  booktitle={Proceedings of the 18th USENIX Symposium on Operating Systems Design and Implementation (OSDI)},
  pages={465--485},
  publisher={{USENIX} Association},
  year={2024}
}

@inproceedings{sel4pm08,
  title={Kernel design for isolation and assurance of physical memory},
  author={Elkaduwe, Dhammika and Derrin, Philip and Elphinstone, Kevin},
  booktitle={Proceedings of the 1st Workshop on Isolation and Integration in Embedded Systems (IIES)},
  pages={35--40},
  publisher={{ACM}},
  doi={10.1145/1435458.1435465},
  year={2008}
}

@inproceedings{sel4fm09,
  title={{seL4}: Formal verification of an {OS} kernel},
  author={Klein, Gerwin and Elphinstone, Kevin and Heiser, Gernot and Andronick, June and Cock, David and Derrin, Philip and Elkaduwe, Dhammika and Engelhardt, Kai and Kolanski, Rafal and Norrish, Michael and Sewell, Thomas and Tuch, Harvey and Winwood, Simon},
  booktitle={Proceedings of the ACM SIGOPS 22nd Symposium on Operating Systems Principles (SOSP)},
  pages={207--220},
  publisher={{ACM}},
  doi={10.1145/1629575.1629596},
  year={2009}
}

@inproceedings{cortenMM25,
    author = {Zhang, Junyang and Xu, Xiangcan and Zou, Yonghao and Tang, Zhe and Wan, Xinyi and Hu, Kang and Wang, Siyuan and Xu, Wenbo and Wang, Di and Chen, Hao and Huang, Lin and Yan, Shoumeng and Tamir, Yuval and Luo, Yingwei and Wang, Xiaolin and Yu, Huashan and Wang, Zhenlin and Tian, Hongliang and Zhou, Diyu},
    title = {{CortenMM}: Efficient Memory Management with Strong Correctness Guarantees},
    year = {2025},
    isbn = {9798400718700},
    publisher = {Association for Computing Machinery},
    address = {New York, NY, USA},
    url = {https://doi.org/10.1145/3731569.3764836},
    doi = {10.1145/3731569.3764836},
    booktitle = {Proceedings of the ACM SIGOPS 31st Symposium on Operating Systems Principles},
    pages = {1082–1098},
    numpages = {17},
    location = {Lotte Hotel World, Seoul, Republic of Korea},
    series = {SOSP '25}
}

@Book{Feske:genode,
  author =       {Norman Feske},
  title =        {Genode Foundations},
  year =         2015,
  publisher =    {Genode Labs},
  url =          {https://genode.org/documentation/genode-foundations/}
}

@inproceedings{metteagle25,
  title={{MettEagle}: Costs and Benefits of Implementing Containers on Microkernels},
  author={Miemietz, Till and Reusch, Viktor and Hille, Matthias and Wrenger, Lars and Eisoldt, Jana and Kl{\"o}tzke, Jan and Kurze, Max and Lackorzynski, Adam and Roitzsch, Michael and H{\"a}rtig, Hermann},
  booktitle={Proceedings of the 19th USENIX Symposium on Operating Systems Design and Implementation (OSDI)},
  pages={979--996},
  publisher={{USENIX} Association},
  year={2025}
}

@inproceedings{m3codesign16,
    author={Asmussen, Nils and V\"{o}lp, Marcus and N\"{o}then, Benedikt and H\"{a}rtig, Hermann and Fettweis, Gerhard},
    title = {{M3}: A Hardware/Operating-System Co-Design to Tame Heterogeneous Manycores},
    year = {2016},
    isbn = {9781450340915},
    publisher = {Association for Computing Machinery},
    address = {New York, NY, USA},
    url = {https://doi.org/10.1145/2872362.2872371},
    doi = {10.1145/2872362.2872371},
    booktitle = {Proceedings of the Twenty-First International Conference on Architectural Support for Programming Languages and Operating Systems},
    pages = {189–203},
    numpages = {15},
    location = {Atlanta, Georgia, USA},
    series = {ASPLOS '16}
}

@misc{ukunfit24,
  author={Cantrill, Bryan},
  title={Unikernels are unfit for production},
  howpublished={\url{https://www.tritondatacenter.com/blog/unikernels-are-unfit-for-production}},
  year={2024}
}

@misc{ukunfit25,
  author={{Unikraft Project}},
  title={Multiprocess support on {Unikraft}},
  howpublished={\url{https://unikraft.org/blog/2025-05-15-multiprocess}},
  year={2025}
}

@inproceedings{mufork25,
  title={$\mu$Fork: Supporting {POSIX} fork Within a Single-Address-Space {OS}},
  author={Kressel, John Alistair and Lefeuvre, Hugo and Olivier, Pierre},
  booktitle={Proceedings of the ACM SIGOPS 31st Symposium on Operating Systems Principles (SOSP)},
  pages={18--35},
  publisher={{ACM}},
  doi={10.1145/3731569.3764809},
  year={2025}
}

@misc{seL42025libos,
  author={{seL4 Foundation}},
  title={{seL4}\_libs: No-assurance Libraries for Rapid-prototyping of {seL4} Applications},
  howpublished={\url{https://github.com/seL4/seL4_libs}},
  year={2025}
}

@book{knuth97,
  title={The Art of Computer Programming: Volume 1: Fundamental Algorithms},
  edition={3rd},
  author={Knuth, Donald E},
  year={1997},
  publisher={Addison-Wesley}
}

@inproceedings{Schatzberg_CDKA_16,
  title={{EbbRT}: A Framework for Building Per-Application Library Operating Systems},
  author={Schatzberg, Dan and Cadden, James and Dong, Han and Krieger, Orran and Appavoo, Jonathan},
  booktitle={Proceedings of the 12th USENIX Symposium on Operating Systems Design and Implementation (OSDI)},
  pages={671--688},
  publisher={{USENIX} Association},
  year={2016}
}

@inproceedings{unikraft21,
  title={Unikraft: fast, specialized unikernels the easy way},
  author={Kuenzer, Simon and B{\u{a}}doiu, Vlad-Andrei and Lefeuvre, Hugo and Santhanam, Sharan and Jung, Alexander and Gain, Gaulthier and Soldani, Cyril and Lupu, Costin and Teodorescu, {\c{S}}tefan and R{\u{a}}ducanu, Costi and Banu, Cristian and Mathy, Laurent and Deaconescu, R\u{a}zvan and Raiciu, Costin and Huici, Felipe},
  booktitle={Proceedings of the 16th European Conference on Computer Systems (EuroSys)},
  pages={376--394},
  publisher={{ACM}},
  doi={10.1145/3447786.3456248},
  year={2021}
}

@phdthesis{sel4mm-phd10,
  title={A principled approach to kernel memory management},
  author={Elkaduwe, Dhammika},
  year={2010},
  doi={10.26190/UNSWORKS/23083},
  school={University of New South Wales},
  address={Sydney, Australia}
}

@misc{barrelbg21,
  author={Baumann, Andrew},
  title={Fish in a {Barrel}: an insider’s retrospective of the {SOSP}'09 multikernel paper},
  howpublished={\url{https://www.sigops.org/2021/fish-in-a-barrel-an-insiders-retrospective-of-the-sosp09-multikernel-paper/}},
  year={2021},
  publisher={{ACM SIGOPS}}
}

@inproceedings{Lampson_71,
  author={Lampson, Butler W.},
  title={Protection},
  booktitle={Proceedings of the 5th Princeton Symposium on Information Sciences and Systems},
  year={1971},
  pages={437--443},
  note={Reprinted in \textit{Operating Systems Review}, 8(1):18--24, 1974}
}

@inproceedings{Boyton_09,
  author={Boyton, Andrew},
  title={A verified shared capability model},
  booktitle={Proceedings of the 4th International Workshop on Systems Software Verification (SSV)},
  pages={25--44},
  publisher={Elsevier},
  year={2009},
  doi={10.1016/J.ENTCS.2009.09.058},
}

@inproceedings{Young_TRGECBBB_87,
  title={The duality of memory and communication in the implementation of a multiprocessor operating system},
  author={Young, Michael and Tevanian, Avadis and Rashid, Richard and Golub, David and Eppinger, Jeffrey and Chew, Jonathan and Bolosky, William and Black, David and Baron, Robert},
  booktitle={Proceedings of the 11th ACM Symposium on Operating System Principles (SOSP)},
  pages={63--76},
  publisher={{ACM}},
  doi={10.1145/41457.37507},
  year={1987}
}

@article{nucleus70,
  title={The nucleus of a multiprogramming system},
  author={Hansen, Per Brinch},
  journal={Communications of the ACM},
  volume={13},
  number={4},
  pages={238--241},
  year={1970},
  doi={10.1145/362258.362278},
  publisher={{ACM}}
}

@article{liedtke1996toward,
  title={Toward real microkernels},
  author={Liedtke, Jochen},
  journal={Communications of the ACM},
  volume={39},
  number={9},
  pages={70--77},
  year={1996},
  publisher={{ACM}}
}

@mastersthesis{weigand2004generalized,
  author={Weigand, Andreas},
  title={Generalized-{Mapping} {IPC} for {L4}}, 
  school={Technische Universit{\"a}t Dresden}, 
  year={2004}, 
  address={Germany}
}

@inproceedings{multikernel09,
  title={The multikernel: a new {OS} architecture for scalable multicore systems},
  author={Baumann, Andrew and Barham, Paul and Dagand, Pierre-Evariste and Harris, Tim and Isaacs, Rebecca and Peter, Simon and Roscoe, Timothy and Sch{\"u}pbach, Adrian and Singhania, Akhilesh},
  booktitle={Proceedings of the ACM SIGOPS 22nd Symposium on Operating Systems Principles (SOSP)},
  pages={29--44},
  publisher={{ACM}},
  doi={10.1145/1629575.1629579},
  year={2009}
}

@inproceedings{Peng_TZLCJXWXZLYZ_25,
  title={{ASTERINAS}: A {Linux} {ABI}-Compatible, {Rust}-Based Framekernel {OS} with a Small and Sound {TCB}},
  author={Peng, Yuke and Tian, Hongliang and Junyang, Zhang and Li, Ruihan and Chen, Chengjun and Jiang, Jianfeng and Xian, Jinyi and Wang, Xiaolin and Xu, Chenren and Zhou, Diyu and Luo, Yingwei and Yan, Shoumeng and Zhang, Yinqian},
  booktitle={Proceedings of the 2025 USENIX Annual Technical Conference (USENIX ATC)},
  pages={307--323},
  publisher={{USENIX} Association},
  year={2025}
}

@inproceedings{unikernel13,
  title={Unikernels: Library operating systems for the cloud},
  author={Madhavapeddy, Anil and Mortier, Richard and Rotsos, Charalampos and Scott, David and Singh, Balraj and Gazagnaire, Thomas and Smith, Steven and Hand, Steven and Crowcroft, Jon},
  booktitle = {Proceedings of the 18th International Conference on Architectural Support for Programming Languages and Operating Systems (ASPLOS)},
  pages={461--472},
  year={2013},
  doi={10.1145/2451116.2451167},
  publisher={{ACM}}
}

@inproceedings{Wrenger_RHDL_23,
  title={{LLFree}: Scalable and Optionally-Persistent Page-Frame Allocation},
  author={Wrenger, Lars and Rommel, Florian and Halbuer, Alexander and Dietrich, Christian and Lohmann, Daniel},
  booktitle={Proceedings of the 2023 USENIX Annual Technical Conference (USENIX ATC)},
  pages={897--914},
  publishe={{USENIX} Association},
  year={2023}
}

@article{eftime2017optimizing,
  title={Optimizing Memory Usage in {L4}-Based Microkernel},
  author={Eftime, Petre and Mogo{\c{s}}anu, Lucian and Caraba{\c{s}}, Mihai and Deaconescu, R{\u{a}}zvan and Gheorghe, Laura and Voiculescu, Valentin Gabriel},
  journal={Computing and Informatics},
  volume={36},
  number={4},
  pages={925--949},
  year={2017},
  doi={10.4149/cai\_2017\_4\_925}
}

@article{knowlton1965fast,
  title={A fast storage allocator},
  author={Knowlton, Kenneth C},
  journal={Communications of the ACM},
  volume={8},
  number={10},
  pages={623--624},
  year={1965},
  publisher={{ACM}},
  doi={10.1145/365628.365655},
}

@article{hirschberg1973class,
  title={A class of dynamic memory allocation algorithms},
  author={Hirschberg, Daniel S},
  journal={Communications of the ACM},
  volume={16},
  number={10},
  pages={615--618},
  year={1973},
  publisher={{ACM}},
  doi={10.1145/362375.362392}
}

@article{burton1976buddy,
  title={A buddy system variation for disk storage allocation},
  author={Burton, Warren},
  journal={Communications of the ACM},
  volume={19},
  number={7},
  pages={416--417},
  year={1976},
  publisher={{ACM}},
  doi={10.1145/360248.360259}
}

@article{peterson1977buddy,
  title={Buddy systems},
  author={Peterson, James L and Norman, Theodore A},
  journal={Communications of the ACM},
  volume={20},
  number={6},
  pages={421--431},
  year={1977},
  publisher={{ACM}},
  doi={10.1145/359605.359626}
}

@article{shen1974weighted,
  title={A weighted buddy method for dynamic storage allocation},
  author={Shen, Kenneth K and Peterson, James L},
  journal={Communications of the ACM},
  volume={17},
  number={10},
  pages={558--562},
  year={1974},
  publisher={{ACM}},
  doi={10.1145/355620.361164}
}

@article{page1986improving,
  title={Improving the performance of buddy systems},
  author={Page, Ivor P and Hagins, Jeff},
  journal={IEEE Transactions on Computers},
  volume={35},
  number={5},
  pages={441--447},
  year={1986},
  publisher={{IEEE}},
  doi={10.1109/TC.1986.1676786}
}

@techreport{wilson1995memory,
  title={Memory Allocation Policies Reconsidered: Methodological Problems and Empirical Realities},
  author={Wilson, Paul R and Johnstone, Mark S and Boles, David and Neely, Michael},
  type={Technical Report},
  institution={Department of Computer Sciences, University of Texas at Austin},
  year={1995}
}

@article{park2014ibuddy,
  title={{iBuddy}: Inverse buddy for enhancing memory allocation/deallocation performance on multi-core systems},
  author={Park, Heekwon and Choi, Jongmoo and Lee, Donghee and Noh, Sam H},
  journal={IEEE Transactions on Computers},
  volume={64},
  number={3},
  pages={720--732},
  year={2015},
  publisher={{IEEE}},
  doi={10.1109/TC.2013.2296049}
}

@article{marotta2021nbbs,
  title={{NBBS}: A non-blocking buddy system for multi-core machines},
  author={Marotta, Romolo and Ianni, Mauro and Pellegrini, Alessandro and Quaglia, Francesco},
  journal={IEEE Transactions on Computers},
  volume={71},
  number={3},
  pages={599--612},
  year={2022},
  publisher={{IEEE}},
  doi={10.1109/TC.2021.3060393}
}

@inproceedings{lu2021spring,
  title={{Spring Buddy}: A Self-Adaptive Elastic Memory Management Scheme for Efficient Concurrent Allocation/Deallocation in Cloud Computing Systems},
  author={Lu, Yihui and Liu, Weidong and Wu, Chentao and Wang, Jia and Gao, Xiaoming and Li, Jie and Guo, Minyi},
  booktitle={Proceedings of the IEEE 27th International Conference on Parallel and Distributed Systems (ICPADS)},
  pages={402--409},
  year={2021},
  publisher={{IEEE}},
  doi={10.1109/ICPADS53394.2021.00056}
}

@misc{google2025cantripos,
  author={Tate-Gans, June and Leffler, Sam and Yick, Kai},
  title={{CantripOS}: An {OS} for {Ambient ML} Applications}, 
  year={2023},
  note={Keynote at the {seL4} Summit 2023},
  howpublished={\url{https://sel4.systems/Foundation/Summit/2023/slides/day2-01-leffler.pdf}},
}

@misc{ramspeed,
  title={{RAMspeed benchmark}},
  author={Hollander, Rhett M and Bolotoff, Paul V},
  howpublished={\url{http://alasir.com/software/ramspeed}},    
  year={2002}
}

@inproceedings{Peter_PVNDS_16,
  title={Undermining isolation through covert channels in the {Fiasco.OC} microkernel},
  author={Peter, Michael and Petschick, Matthias and Vetter, Julian and Nordholz, Jan and Danisevskis, Janis and Seifert, Jean{-}Pierre},
  booktitle={Proceedings of the 30th International Symposium on Computer and Information Sciences (ISCIS)},
  pages={147--156},
  year={2015},
  doi={10.1007/978-3-319-22635-4\_13},
  publisher={Springer}
}

@inproceedings{Nelson_SZJBTW_17,
  title={Hyperkernel: Push-button verification of an {OS} kernel},
  author={Nelson, Luke and Sigurbjarnarson, Helgi and Zhang, Kaiyuan and Johnson, Dylan and Bornholt, James and Torlak, Emina and Wang, Xi},
  booktitle={Proceedings of the 26th Symposium on Operating Systems Principles (SOSP)},
  pages={252--269},
  doi={10.1145/3132747.3132748},
  publisher={{ACM}},
  year={2017}
}

@article{Hartig_HLSW_97,
  title={The performance of $\mu$-kernel-based systems},
  author={H{\"a}rtig, Hermann and Hohmuth, Michael and Liedtke, Jochen and Sch{\"o}nberg, Sebastian and Wolter, Jean},
  journal={Proceedings of the 16th ACM Symposium on Operating System Principles (SOSP)},
  pages={66--77},
  year={1997},
  doi={10.1145/268998.266660},
  publisher={{ACM}}
}

@misc{SQLite,
  title={{SQLite}},
  author={SQLite Consortium},
  howpublished={\url{https://www.sqlite.org/index.html}},
  year={2018}
}

@inproceedings{ycsb,
  title={{Benchmarking cloud serving systems with YCSB}},
  author={Cooper, Brian F and Silberstein, Adam and Tam, Erwin and Ramakrishnan, Raghu and Sears, Russell},
  booktitle={Proceedings of the 1st ACM symposium on Cloud computing},
  pages={143--154},
  doi={{10.1145/1807128.1807152}},
  publisher={{ACM}},
  year={2010}
}

@inproceedings{xv6fs,
  title={Using Crash Hoare logic for certifying the {FSCQ} file system},
  author={Chen, Haogang and Ziegler, Daniel and Chajed, Tej and Chlipala, Adam and Kaashoek, M Frans and Zeldovich, Nickolai},
  booktitle={Proceedings of the 25th Symposium on Operating Systems Principles},
  pages={18--37},
  doi={10.1145/2815400.2815402},
  publisher={{ACM}},
  year={2015}
}

@misc{seL4_hw_support,
  author={{seL4 Foundation}},
  title={Supported platforms of the {seL4} microkernel}, 
  year={2025},
  howpublished={\url{https://docs.sel4.systems/Hardware}},
}

@inproceedings{Szekely_BMK_24,
  title={Unifying serverless and microservice workloads with {SigmaOS}},
  author={Szekely, Ariel and Belay, Adam and Morris, Robert and Kaashoek, M Frans},
  booktitle={Proceedings of the ACM SIGOPS 30th Symposium on Operating Systems Principles (SOSP)},
  pages={385--402},
  publisher={{ACM}},
  doi={10.1145/3694715.3695947},
  year={2024}
}

@misc{linux_hugetlb,
  author={{Linux Kernel Community}},
  title={{HugeTLB} Pages},
  howpublished={\url{https://docs.kernel.org/admin-guide/mm/hugetlbpage.html}}
}

@misc{eembc_coremark,
  author       = {{EEMBC}},
  title        = {{CoreMark}},
  url          = {https://www.eembc.org/coremark/},
  note         = {Accessed: 2026-10-09}
}

@inproceedings{Peters_DEH_15,
  title={For a Microkernel, a Big Lock Is Fine},
  author={Peters, Sean and Danis, Adrian and Elphinstone, Kevin and Heiser, Gernot},
  booktitle={Proceedings of the 6th Asia-Pacific Workshop on Systems (APSys)},
  numpages={7},
  publisher={{ACM}},
  doi={10.1145/2797022.2797042},
  year={2015}
}

@article{sel4comp14,
  title={Comprehensive formal verification of an {OS} microkernel},
  author={Klein, Gerwin and Andronick, June and Elphinstone, Kevin and Murray, Toby and Sewell, Thomas and Kolanski, Rafal and Heiser, Gernot},
  journal={ACM Transactions on Computer Systems (TOCS)},
  volume={32},
  number={1},
  pages={2:1--2:70},
  year={2014},
  doi={10.1145/2560537},
  publisher={{ACM}}
}

@misc{seL42025microkit,
  author={{seL4 Foundation}},
  title={The {seL4} Microkit},
  howpublished={\url{https://docs.sel4.systems/projects/microkit/}},
  year={2025}
}

@manual{sel4manual13,
  title        = {seL4 Reference Manual},
  author       = {{seL4 Foundation}},
  year         = {2025},
  version      = {13.0.0},
  note         = {Available at \url{https://docs.sel4.systems/}}
}

\end{document}